\documentclass[a4paper,11pt]{article}
\pdfoutput=1 

\usepackage{jcappub} 

\usepackage[T1]{fontenc} 
\usepackage{hyperref}
\usepackage{orcidlink}
\usepackage{lineno}
\usepackage{multirow}
\usepackage{tikz}

\title{
\boldmath Neutrino mass ordering from the next  Galactic supernova  at DUNE, HK, and JUNO\\
}

\author[a]{Prantik Sarmah\orcidlink{0000-0003-3159-7148}}
\author[b]{, Sovan Chakraborty\orcidlink{0000-0002-1458-8517}}
\author[b]{, Abinash Medhi\orcidlink{0000-0002-1518-4354}}
\author[c]{, Debanjan Bose\orcidlink{0000-0003-1071-5854}}
\author[d]{, and  Moon Moon Devi\orcidlink{0000-0001-6569-0792}}

\affiliation[a]{State Key Laboratory of Particle Astrophysics, Institute of High Energy Physics, Chinese Academy of Sciences, Beijing, 100049,  China}
\affiliation[b]{Department of Physics, Indian Institute of Technology Guwahati, North Guwahati- 781039, India}
\affiliation[c]{Department of Physics, Central University of Kashmir, Main Campus, Tulmulla, Ganderbal- 191131, Jammu and Kashmir, India}
\affiliation[d]{Department of Physics, Tezpur University, Napaam, Assam, India}

\emailAdd{prantiksarmah@ihep.ac.cn}
\emailAdd{amedhi0@rnd.iitg.ac.in}
\emailAdd{debanjan.tifr@gmail.com}
\emailAdd{sovan@iitg.ac.in}
\emailAdd{devimm@tezu.ernet.in}

\abstract{The next Galactic core-collapse supernova (CCSN) will offer a unique opportunity to determine the neutrino mass ordering. 
We focus on two observables: the electron neutrino ($\nu_e$) neutronization burst and the rise-time of the electron antineutrino ($\bar{\nu}_e$) flux during the accretion phase. The neutronization burst, a sharp $\nu_e$ peak within $\sim 20$--$30$ ms, provides a clean and robust signature of mass ordering through its appearance 
or disappearance.  
During the accretion phase, the faster rise of heavy lepton flavor neutrinos ($\nu_x$) leads to a distinct faster rise-time behavior of the oscillated $\bar{\nu}_e$ signal, resulting in mass ordering discrimination.
Using realistic CCSN simulations for multiple progenitor masses, we compute event rates and perform a statistical analysis for a Galactic ($10$~kpc) CCSN event at DUNE, Hyper-Kamiokande (HK), and JUNO detectors. The neutronization burst remains largely independent of SN hydrodynamic simulation models, with DUNE and HK achieving $\gtrsim 6\sigma$ and $\gtrsim 4\sigma$ sensitivity for normal (NO) to inverted ordering (IO) discrimination, respectively. However, the rise-time observable is prone to  progenitor degeneracies. To mitigate this 
cumulative and ratio-based observables constructed at characteristic timescales ($20$ ms \& $100$ ms) are used. 
The resulting confidence levels from the rise-time analysis to discriminate IO/NO in HK and JUNO are $\sim 5\sigma$ and $\sim 3\sigma$, respectively.  
Our results highlight the complementarity of detectors and observables, and demonstrate that combining neutronization burst and accretion phase information will be crucial for a definitive determination of the neutrino mass ordering in the next Galactic supernova.
}

\begin{document}
\maketitle
\flushbottom

\section{Introduction}
\label{sec:intro}

A Galactic supernova (SN) explosion  will be a pivotal event providing not only the understanding of the physics of SN explosions but also the fundamental properties of the neutrinos such as the mass ordering~\cite{Scholberg:2017czd}. This is because supernova explosions are expected to create a huge amount of MeV neutrinos, $\mathcal{O}(10^{58})$, giving rise to a large number of events at the different Earth-based neutrino detectors.  The number of neutrinos detected from SN 1987A was only about $20$ events across all detectors, primarily due to the limited detector volumes available at that time \cite{Kamiokande-II:1987idp,1987PhRvL..58.1494B,Alekseev:1987ej}. The capabilities of the present neutrino detectors such as Super-kamiokande (SK) and JUNO  to study a Galactic supernova event have improved significantly. Given the rate of SN explosion in the Milky Way to be about $1$-$2$ per century~\cite{Rozwadowska:2020nab,10.1093/mnras/stab2182,1991ARA&A..29..363V}, the next Galactic SN is expected to light up the sky anytime soon. Thus, in this work, we explore the capabilities of detectors such as Deep Underground Neutrino Experiment (DUNE)~\cite{DUNE:2015lol}, Hyper-Kamiokande (HK)~\cite{Hyper-Kamiokande:2018ofw}, and  Jiangmen Underground Neutrino Observatory (JUNO)~\cite{JUNO:2015zny} to probe the neutrino mass ordering using  statistical analysis of SN neutrinos from the early emission phases.

During a Core collapse SN (CCSN) explosion, neutrinos of all flavors are copiously produced through a variety of microphysical processes that operate at different stages following the core bounce~\cite{Mirizzi:2015eza,Mueller:2026acg,Tamborra:2014hga,Nakazato,Vartanyan:2021dmy}. The post-bounce evolution is typically divided into three main phases:  \textit{the neutronization burst}, \textit{the accretion}, and \textit{the cooling}, each characterized by distinct emission mechanisms and neutrino signatures. The durations of these phases are typically of the order of $\sim 30$~ms, a few  $100$~ms, and a few  $10$s of seconds, respectively.


The neutrino emission during the neutronization burst and accretion phases exhibits strong flavor hierarchies, making these phases highly sensitive to flavor oscillation effects induced during propagation~\cite{Kachelriess:2004ds,Dighe:1999bi,DUNE:2020zfm,Mirizzi:2015eza,Panda:2023rxa,Brdar:2022vfr}. The neutronization burst is composed almost entirely of a sharp peak of electron neutrinos ($\nu_e$) produced via rapid electron capture on protons.
The presence or suppression of this peak at detectors has been proposed as a signature of the true neutrino mass ordering; in particular, the peak is strongly suppressed for normal ordering (NO, $\Delta m_{\rm atm}^2 > 0$)~\cite{Dighe:1999bi}. During the accretion phase, neutrinos of all flavors are abundantly produced via processes such as electron and positron captures along with thermal pair processes. However, as the core gradually deleptonizes, the luminosities of the heavy-lepton flavors ($\nu_\mu$, $\nu_\tau$ and their antiparticles) increase more rapidly than those of the electron flavors, leading to pronounced flavor hierarchies in the emitted neutrino spectra.
This  faster rise of the heavy-lepton flavor fluxes relative to the electron antineutrino flux is then altered by oscillations during propagation from the SN core to the outer stellar surface, causing the observed electron antineutrino flux to rise more rapidly for inverted ordering (IO, $\Delta m_{\rm atm}^2 < 0$) than for NO~\cite{Serpico:2011ir}. Consequently, the rise time of the accretion-phase electron antineutrino flux has been proposed as a probe of the neutrino mass ordering. The only existing study in the literature for this rise time proposal is
based on the IceCube detector ~\cite{Serpico:2011ir}.  Thus, a detailed analysis for detectors such as HK and JUNO, for rise time together with the neutronization burst phenomenology at HK and DUNE becomes crucial. Therefore, in this work, we perform a statistically robust analysis of the neutrino mass ordering for these three detectors for a Galactic SN ($10$ kpc).

These proposals are also not without challenges; in particular, there are complications arising from the complex hydrodynamics of the dense SN core that influence our understanding of the initial neutrino flux diffusing out of their last scattering surface, i.e., the neutrinosphere ~\cite{Janka:2025tvf,Mezzacappa:2020oyq,Mirizzi:2015eza}. In addition, the flavor oscillation of these diffusing neutrinos deep inside the star, close to the neutrinosphere, can induce collective neutrino flavor conversions ($\sim 10^1-10^2$~km) ~\cite{Pantaleone:1992eq,Samuel:1993uw,Samuel:1996ri,Pastor:2001iu,Pastor:2002we,Duan:2006an,Duan:2010bg} in addition to the standard Mikheyev–Smirnov–Wolfenstein (MSW, few $\sim 10^3$~km) effects~\cite{Dighe:1999bi}. These collective oscillations are generally classified into two types: vacuum-driven modes and fast modes triggered by the swaps in the energy and angular spectra of neutrinos, respectively. These collective effects are proposed to be fast and may take place on the scale of few $cm$ to $m$, opening the possibility of fast flavor conversions and equal ratio of different neutrino flavors \cite{Capozzi:2018rzl}.
However, the present understanding of these effects is still based on several approximations, lacking a clear final picture~\cite{Sen:2024fxa,Tamborra:2020cul}. In particular, we focus on the proposal, where close to the SN core the large matter density over the neutrino one suppresses the growth of the collective modes and only MSW effects remain important. Although we do not explicitly probe these collective oscillation effects, rather take into account the scenario in which any fast growth of the flavor instability may equalize the neutrino flavor ratios upon escape from the supernova core, as proposed in \cite{Capozzi:2018rzl}. Thus, in general, we consider three oscillation scenarios: normal ordering (NO), inverted ordering (IO), and flavor equalization (FE). Note that this FE scenario, with its origin in fast growth of the collective modes, remains  suppressed during the neutronization burst phase due to the high electron degeneracy. 

For the source SN neutrino flux, we consider six different SN simulation models corresponding to different progenitor masses~\cite{Garching,Gonzalez-Garcia:2021dve}. Subsequently, we analyze the expected neutrino events in the aforementioned detectors for  each combination of simulation model and oscillation scenario. 
The distinctive feature of the neutronization burst phase is considered as the robust probe of the true MO independent of the progenitor model, manifested through the appearance (IO) or disappearance (NO) of the burst peak. This phenomenology is particularly pronounced in DUNE and remains reasonably effective in HK. 
For the rise time analysis during the accretion phase, to achieve a similar progenitor model independence of the true MO, we need a robust statistical analysis of the observed neutrino events. 
Given the inherent uncertainties in the progenitor properties and the different methods, there may exist a non-negligible probability of MO misidentification. To quantify this, we perform a statistical analysis for both the neutronization and rise time phases to estimate this misidentification probability.
For the neutronization burst phase, we find that for our choice of the progenitor models the  misidentification probability is less than $10^{-10}$ ($\sim 6~\sigma$ confidence level) and less than $10^{-6}$ ($\sim 4~\sigma$ confidence level) for DUNE and HK, respectively. 
Thus, the MO can be confidently identified at both DUNE and HK for a future Galactic SN event. For the rise time analysis, the misidentification probabilities for HK and JUNO are found to be   $\lesssim 0.01$ ($\sim 3~\sigma$ confidence level) and $\lesssim 0.4$ ($\sim 1~\sigma$ confidence level). These misidentification probabilities are larger compared to those for the neutronization burst analysis because of the FE scenario. As already mentioned, the FE scenario arising from collective oscillations remains an open question and is an active area of ongoing research. Therefore, future advances in the understanding of collective oscillations will enable more accurate constraints on such misidentification probabilities. Interestingly, determining the true neutrino MO from the neutronization burst neutrinos could also provide valuable insights into the nature of collective oscillations.

This paper is organized as follows. In Sec.~\ref{sec:SN_models}, we introduce the SN models considered for this analysis. Sec.~\ref{sec:mass_ordering} comprises discussion of the SN neutrino emission mechanisms, neutrino oscillation in SN and their mass ordering signals. A brief overview of the detectors and event rate calculator (SNOwGLoBES) is provided in Sec.~\ref{sec:detectors}. Sec.~\ref{sec:results} is dedicated to the results of our study which are complemented with a detailed statistical analysis in Sec.~\ref{sec:statistical_analysis}.  The paper is finally concluded in Sec.~\ref{sec:conclusion}.

\begin{figure}[!t]
    \centering
    \includegraphics[width=0.95\textwidth]{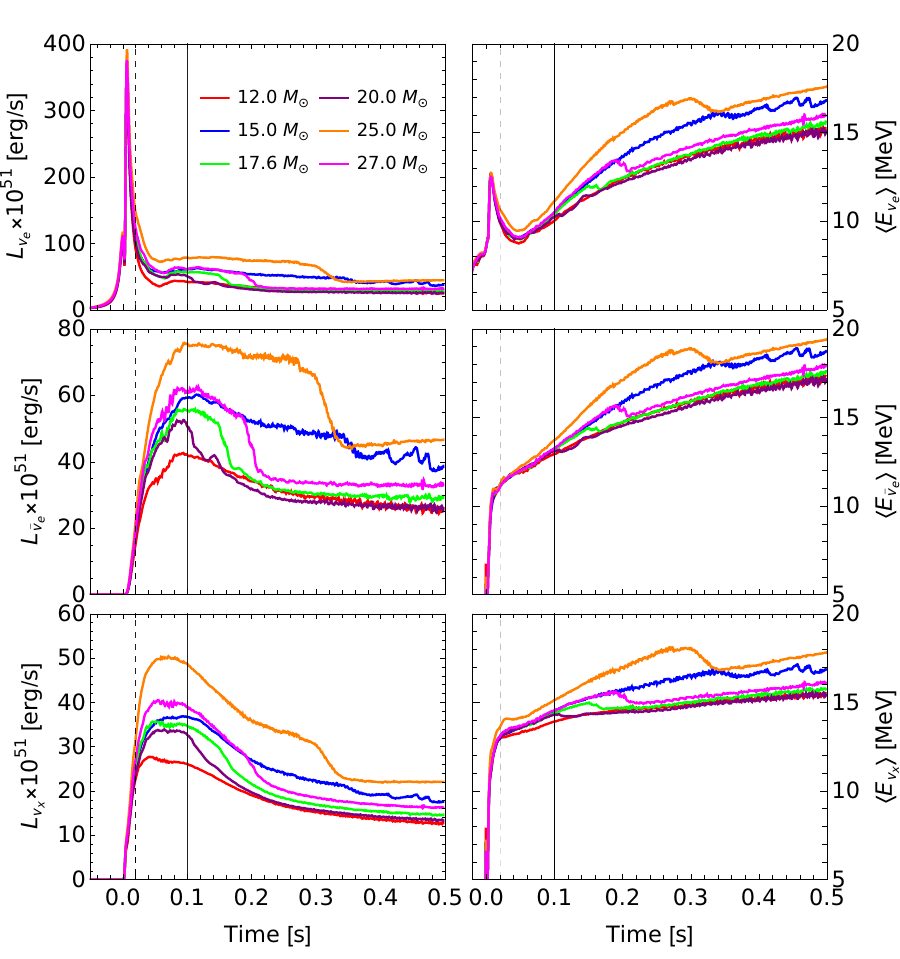}
    \caption{Luminosities (left) and average energies (right) for different neutrinos emitted by CCSNe simulation of different progenitor masses \citep{Garching}. The top panel shows the electron neutrinos and the middle panel is for electron anti-neutrinos. Muon and tau flavored neutrinos and anti-neutrinos are treated as same (labelled as $\nu_{\rm x}$) and shown in the bottom panel. The vertical dashed and continuous lines are drawn at 20 ms and 100 ms respectively. }
    \label{fig:light_curves}
\end{figure}

\section{Supernova models: primary neutrino flux}
\label{sec:SN_models}
Hydrostatic burning of nuclear fuel leads a star to the formation of a degenerate iron or oxygen-neon-magnesium cores that collapses due to strong gravitational pull \cite{Mirizzi:2015eza}. This collapse gives rise to the creation of a huge number of MeV neutrinos$\sim\mathcal{O}(10^{58})$ in about a few seconds, that is possibly accompanied by a supernova explosion. The star might eventually end up as a compact object (black hole or neutron star). The uncertainties of the details of such a CCSN mechanism are crucial to probe any future galactic SN event and demand realistic hydrodynamic simulations to settle the ambiguities. Such realistic simulation of CCSN in the accurate three dimensional framework is still not achievable \cite{Janka:2016fox,Garching}. However, models with reasonable approximations ranging in all three dimensions have yielded  crucial information about the possible properties of the SN neutrinos\cite{Garching,Vartanyan:2021dmy,Shibagaki:2023tmh,Vartanyan:2024bhx,Halevi:2025mga,Bruenn:2009ucj}. These available simulations of CCSN in literature use hydrodynamic codes based on various simplified assumptions to solve the equation of state \cite{LSEOS,Shen:1998gq,Steiner:2012rk}, and these assumptions are different for different simulation models. Thus, the neutrino fluxes for different simulated models show different characteristic features. In this section, we provide a brief overview of the different Garching simulations of CCSN adopted for our analysis~\cite{Garching}. 

All the Garching simulation models used here, are obtained using the PROMETHEUS-VERTEX code as it solves the hydrodynamic  equations of states for neutrino number, energy, and momentum of moments $\mathcal{O}(v/c)$, with a variable Eddington factor closure \cite{Garching}. These simulations are performed using "ray-by-ray plus" method to take into account of two and three dimensional effects assuming azimuthal symmetry around the radial direction. Although the hydrodynamic treatment is Newtonian, a reasonable correction to the gravitational potential has been invoked for the approximate general relativistic effects. 
We use the results of the simulation for the Lattimer and Swesty EOS with a nuclear incompressibility of $220$ MeV. We take six different progenitor mass models, namely the models with the following progenitor mass in Solar mass units $12$, $15$, $17.6$, $20$, $25$ and $27$. These models provide the neutrino luminosity and average energy as  functions of time. The luminosities (left panels) and average energies (right panels) of the different neutrino species for these models are shown in Fig.~\ref{fig:light_curves}. The top, middle, and bottom panels correspond to the $\nu_e$, $\bar{\nu}_e$, and $\nu_x$ species, respectively. The neutrino species exhibit distinct spectral characteristics, rendering them to be clearly distinguished from one another. These differences arise because the production and emission of the various neutrino species are governed by different physical processes operating at different phases of evolution. In the following, we discuss the details of the different neutrino emission phases and corresponding neutrino emission characteristics.




\subsection{SN neutrino emission phases}
\label{sec:mass_ordering}
 The neutrino emission during the first few 100 milliseconds carries crucial information about the possible neutrino mixing \cite{Serpico:2011ir, Kachelriess:2004ds}. 
This first 100 milliseconds consist of two different phases namely, the  {\it neutronization burst}  (about $20-30$~ms post bounce) and the {\it   accretion} (about a second post bounce). These  early neutrinos exhibit  significant spectral hierarchies between different species of neutrinos. 
Thus, any  neutrino flavor oscillation in this early phase will have the most chances for phenomenological detection. 
In the following, we describe in detail these neutrino emission processes, their flavor conversion and possible implementations to probe the neutrino mass ordering. \\

\textbf{Neutronization burst:} In the CCSN model, the collapse of the massive star results in this extremely dense ($\sim 10^{14}~\rm gcm^{-3}$) core that bounces off creating a strong shock propagating radially outward through the dense matter.   This core bounce also results in a prompt burst of electron neutrinos ($\nu_{e}$) that are created by electron capture on protons. Consequently, most of the protons get converted into  neutrons (deleptonization) and hence the name neutronization burst. The burst occurs when the shock reaches density low enough (the so called neutrinosphere) for the neutrinos to escape and lasts for about 20-30 ms since the onset of the core bounce. Note that this duration might vary depending on the specific simulation models \cite{Garching,Tamborra:2014hga,Vartanyan:2021dmy,Nakazato:2012qf}.

This phase is thus characterized by this burst of electron neutrinos, resulting in a large peak in the $\nu_e$ luminosity. 
Electron anti-neutrinos ($\Bar{\nu}_{e}$) and the heavy lepton flavors ($\nu_{x}$) are majorly produced via the electron-positron pair annihilation and the nucleon-nucleon bremsstrahlung. Note that $\Bar{\nu}_{e}$ can also be created by positron capture on neutron.   However, the production of $\bar{\nu_e}$ and the $\nu_{x}$ during this neutronization burst phase is heavily suppressed due to the high electron degeneracy in the core. Therefore, the neutrino luminosity of the neutronization burst is  dominated by a sharp peak in the $\nu_{e}$ spectra as shown in Fig.~\ref{fig:light_curves}.

The average energies of these neutrinos as a function of time are shown in the right panel of Fig.~\ref{fig:light_curves}. Signature of the neutronization burst also appears as a peak in the energies of $\nu_{e}$ (top right). This is because the energies of the electrons increases sharply with the sudden rise in temperature due to core collapse. These energies attain a maximum and falls rapidly due to the core bounce. Both the luminosity and average energy of $\nu_{e}$ falls rapidly after the prompt burst till the point when the deleptonization saturates as the proton density becomes too low for electron capture, resulting in an extremely neutron rich core.  

\textbf{Accretion:} At the end of this deleptonization, the core temperature starts increasing as matter keeps accreting to the core. This rise in the temperature is expected to facilitate electron-positron pair annihilation and nucleon-nucleon bremsstrahlung, resulting in efficient pair production of neutrinos-antineutrino of all flavors including the mu and tau types. These mu and tau type neutrino-antineutrino ($\nu_{x}$) production is a unique characteristic of this accretion phase. Regarding the $\Bar{\nu}_{e}$, it is important to realize that the $\Bar{\nu}_{e}$ production via positron capture will remain suppressed due to the high electron degeneracy in the initial phase.  Subsequently, during this pair production phase $\Bar{\nu}_{e}$ population is also suppressed due to the high abundance of $\nu_{e}$ and the corresponding fermion blocking in the $\nu_{e}$ phase space \cite{Capozzi:2018rzl,Mirizzi:2015eza}. Indeed, once the $\Bar{\nu}_{e}$ becomes dense enough, their luminosity is further reduced by the neutrino anti-neutrino pair annihilation, $\nu_{x}$ ($\nu_{e}\Bar{\nu}_{e} \longrightarrow 2 \nu_{x}$). This makes a faster rise in the luminosity of the $\nu_{x}$, in contrast to that of the $\Bar{\nu}_{e}$ which rises slower.

The faster rise of the luminosities of these $\nu_{x}$s flavors of neutrinos in comparison to that of the $\Bar{\nu}_{e}$ is found to be a general feature of different CCSN simulation models , e.g., see Fig.~\ref{fig:light_curves} and \cite{Garching,Vartanyan:2021dmy,Tamborra:2014hga}. Typically, all the models show a consistent feature that the luminosities of the different neutrino flavors rise to their maxima at around $100$ ms. Post this epoch the neutrino luminosities keep falling gradually (see Fig.~\ref{fig:light_curves} middle and bottom left panels) as the expanding shock prevents matter from reaching the PNS resulting in a decreasing mass accretion rate \cite{Janka:2012,Janka:2012wk}. 

In fact, the neutrino emission and the interaction of the shock with the dense matter eventually slow down the shock. This stalling of the shock may further increase the accretion of matter leading to a failed SN. In case of successful explosion, the shock may revive due to energy deposition by neutrinos to the matter behind the shock as a result of neutrino-matter interaction \cite{Janka:2012wk}. This shock revival may also take place due to magnetorotational effects, hydrodynamic instabilities such as convection and the standing accretion shock instability (SASI), acoustic energy deposition, or jet-driven mechanisms \cite{Janka:2012wk}. Nevertheless, the question of shock revival in successful SN explosions remains unresolved and is not yet fully understood.


The middle and the bottom right panels of Fig.~\ref{fig:light_curves} show the average energies for the accretion phase neutrinos. The average energy of $\nu_{x}$ is generally larger than that of the electron flavor neutrinos. This is because the $\nu_{x}$ decouple from the matter earlier (smaller radius) than that electron flavor neutrinos as they do not have sufficient energies to initiate charged-current interactions with the matter. On the other hand, electron flavored neutrinos (anti-neutrinos) interact with the matter to produce electrons (positrons) and decouple at larger radius. Therefore, the neutrinosphere ($R_\nu$) of $\nu_{x}$ is smaller than that of $\nu_{e}$ and $\Bar{\nu}_{e}$. The neutrinos being in thermal equilibrium with the background their average energies can be described in terms the neutrinosphere radius. Different SN models show this typical behavior with some differences among them due to model details.   
 
Eventually, for any successful supernova explosion scenarios the shock revival is crucial. In such cases the density becomes low enough for neutrinos to diffuse out and cooling of the star begins via this neutrino emission. The luminosities of different neutrino flavors become nearly equal during this phase and might last about up to $20$-$30$ seconds \cite{Nakazato:2012qf,Tamborra:2014hga,Garching}. Note that for the models adopted in this analysis, the simulation data are given up to the accretion phase ($0.5$~s)~\cite{Gonzalez-Reina:2022ehy} and cooling curves are not shown in Fig.~\ref{fig:light_curves}.


Given the luminosities and average energies of the different neutrino species, one can obtain the corresponding fluxes at the source and discussed in the following. Clearly, these fluxes would be different for the different emission phases (neutronization burst and accretion) as the luminosities and the average energies vary with time. 

\subsection{Neutrino Flux}
\label{subsec:neutrino_flux}

The flux of these neutrinos at the source is assumed to be quasi-thermal and modeled as~\cite{Tamborra:2012ac,Serpico:2011ir,Keil:2002in}, 

\begin{equation}
    F^{0}_{\nu}(E_{\nu},t)= L_{\nu}(t) \frac{\left(1+\beta_{\nu}(t)\right)^{1+\beta_{\nu}(t)}}{\Gamma\left(1+\beta_{\nu}(t)\right)} \times \frac{E_{\nu}^{\beta_{\nu}(t)}}{\left(\langle E_{\nu} (t) \rangle\right)^{\beta_{\nu}(t)+2}} \times \exp{\left[-(\beta_{\nu}(t)+1) \frac{E_{\nu}}{\langle E_{\nu}(t) \rangle} \right]} ~,
    \label{eq:source_flux}
\end{equation}

where $\nu$ represents $\nu_{\rm e}$, $\Bar{\nu}_{\rm e}$, $\nu_{\rm x}$ and $\beta_{\nu}(t)$ is energy-shape parameter given by

\begin{equation}
    \beta_{\nu}(t)=\frac{2 (\langle E_{\nu} (t) \rangle)^2-\langle E_{\nu} (t)^2 \rangle}{\langle E_{\nu}(t)^2 \rangle -(\langle E_{\nu} (t) \rangle)^2 }~, 
\end{equation}

where $\langle E_{\nu}(t)^2 \rangle$ is the mean of squared neutrino energy and obtained from the simulation data~\cite{Garching}. 


The neutrinos produced in CCSN might  undergo two different flavor conversion processes during propagation in the dense medium to the outer layer of the collapsed star. The collective oscillations due to interaction among the neutrinos themselves occurs within radius $\mathcal{O}(10^3)$ km, whereas the MSW oscillations become important further out around $\mathcal{O}(10^4)$ km, thus making these two processes independent of each other \cite{Serpico:2011ir}.
The collective oscillations are proposed to grow from two different origins, 
the neutrino mass dependent one typically termed as the `slow' collective modes 
and the neutrino density dependent `fast' oscillations. The typical `slow'  collective oscillations in iron-core SNe are suppressed by trajectory dependent multi-angle effects associated with the dense matter \cite{Chakraborty:2011gd,Chakraborty:2011nf, Sarikas:2011am}.  In fact, the matter density in the core largely exceeds the neutrino density during the early time since the core bounce, i.e, until about 0.2 s. Hence, the slow collective oscillations is expected to be strongly suppressed during this period \cite{Chakraborty:2011gd,Chakraborty:2011nf,Sarikas:2011am}. However, self-induced neutrino fast flavor conversion might still be possible in the SN core \cite{Capozzi:2018rzl}. This effect is possible due to the high neutrino density in the SN core and, in the extreme scenario, this may equalize flavor ratio and may wash out any further oscillation impact.  In fact, this scenario of flavor equilibration (FE) should be able to probe other proposals where the slow oscillations may grow bit faster than usual, though slower than the fast ones \cite{Fiorillo:2025gkw, Fiorillo:2026tee}. However, these fast oscillations require a specific scenario in which  the neutrino and antineutrino angular distributions have at least one crossing. Whether, such  crossings exist in the emitted neutrinos or not depends on the specifics of the SN simulations and the emission direction of these neutrinos. The crossings have been seen to originate at different distances as a result of different processes, such as collisions and backward scattering. The understanding of these fast oscillations is still growing and  not settled, hence we treat the FE scenario as an effective signal for any such oscillations beyond MSW. 
Interestingly, collective oscillations, i.e., the FE scenario remain suppressed in the neutronization burst phase due to high electron number degeneracy, and might become important only during the accretion phase.  Given the fluxes of different flavor at production ($F_{\nu_{\rm e}}^0$, $F_{\Bar{\nu}_{\rm e}}^0$  and $F_{\nu_{\rm x}}^0$) in Eq.~\ref{eq:source_flux}, the oscillated neutrino fluxes ($F_{\nu_{\rm e}}$, $F_{\Bar{\nu}_{\rm e}}$ and $F_{\nu_{\rm x}}$) for different scenarios are given below.  The NO and IO cases are solely due to MSW with matter suppression of collective oscillations \cite{Chakraborty:2012gb,Serpico:2011ir,Dighe:1999bi}; thus similar for both neutronization and accretion phase, whereas the FE scenario is for accretion only \cite{Capozzi:2018rzl},\\

{\bf Normal ordering (NO):}
\begin{align}
    F_{\nu_{\rm e}} &= F_{\nu_{\rm x}}^0~, \label{eq:NO_nue} \\
    F_{\bar{\nu}_{\rm e}} &= \cos^2{\mathcal{\theta}_{12}}
    F_{\bar{\nu}_{\rm e}}^0 + \sin^2{\mathcal{\theta}_{12}} F_{\nu_{\rm x}}^0~.
    \label{eq:NO_nubare}
\end{align}

{\bf Inverted ordering (IO):} 
\begin{align}
F_{\nu_{\rm e}} &= \sin^2{\mathcal{\theta}_{12}} F_{\nu_{\rm e}}^0 + \cos^2{\mathcal{\theta}_{12}} F_{\nu_{\rm x}}^0~,  \label{eq:IO_nue}  \\
    F_{\bar{\nu}_{\rm e}} &= F_{\nu_{\rm x}}^0~. \label{eq:IO_nubare}
\end{align}


{\bf Flavor equalisation (FE):} 
\begin{align}
F_{\nu_{\rm e}} &= \frac{1}{3} F_{\nu_{\rm e}}^0 + \frac{2}{3} F_{\nu_{\rm x}}^0~,  \label{eq:FE_nue}  \\
    F_{\Bar{\nu}_{\rm e}} &=  \frac{1}{3} F_{\Bar{\nu}_{\rm e}}^0  + \frac{2}{3} F_{\nu_{\rm x}}^0~. \label{eq:FE_nubare}
\end{align}

For the solar mixing angle $\mathcal{\theta}_{12}$ we take $\sin^2{\mathcal{\theta}_{12}} = 0.31$ \cite{Esteban:2024qld}. 
The features of the neutronization burst and the rise of $\nu_{x}$ discussed above can be traced using the above equations. For example, $F_{\nu_{e}}$ will show the neutronization burst peak in IO as this flux contains about 31 \% of the unoscillated $\nu_{e}$ flux, $F_{\nu_{e}}^0$. On the other hand, this peak should be absent for NO as in this case all $\nu_{e}$ comes from only $\nu_{x}$. Note that we are interested in the fluxes, $F_{\nu_{e}}$ and $F_{\Bar{\nu}_{e}}$ only, as we aim to probe the oscillation scenarios at neutrino detectors sensitive to the electron flavor neutrinos and antineutrinos. We briefly discuss  the relevant detectors in the next section. 

\section{Detectors}
\label{sec:detectors}
In order to study the mass ordering phenomenologies of the neutronization burst and rise time, we focus on detectors capable of detecting electron neutrinos and electron anti-neutrinos, respectively. For the neutronization burst, we consider two upcoming detector set up, Hyper Kamiokande (HK) and DUNE, whereas for the rise time, we take HK and the currently running JUNO detectors. 
In the following, we discuss the details of the three upcoming detectors used in our analysis.

\textit{DUNE:}
The Deep Underground Neutrino Experiment (DUNE) \cite{DUNE1,DUNE2,DUNE3,DUNE4,DUNE5} is a proposed long-baseline superbeam neutrino experiment with a baseline of 1300 km. The primary scientific objectives of this experiment include: (a) determining the neutrino mass ordering, the octant of $\theta_{23}$, and the presence of CP violation in the leptonic sector using accelerator neutrinos; and (b) detecting neutrinos from core-collapse supernovae~\cite{Ankowski:2016lab,DUNE:2020zfm}.
SN neutrinos can be detected in DUNE using the far detector, which is a Liquid Argon Time Projection Chamber (LArTPC) consisting of four identical detectors with a total fiducial mass of 40 kt. 
The primary detection channel in this detector is the charged-current interaction of $\nu_e$ with argon, $\nu_e + {}^{40}\mathrm{Ar} \rightarrow e^- + {}^{40}\mathrm{K}^\star$.
Consequently, this detector is particularly sensitive to the electron neutrinos produced during the neutronization burst phase of a core-collapse supernova. In addition, neutrinos and antineutrinos of all flavors can also be detected through neutral-current interactions, $\nu_X + \mathrm{Ar} \rightarrow \nu_X + \mathrm{Ar}^\star $, where $X$ denotes any neutrino flavor.


\textit{Hyper-Kamiokande:} Hyper-Kamiokande (HK) is a next-generation water Cherenkov detector currently under construction in Hida, Japan~\cite{Hyper-Kamiokande:2018ofw}. 
This detector will have a large fiducial mass of 187 kt. 
In addition to the detection of SN neutrinos~\cite{Hyper-Kamiokande:2021frf},  it has diverse physics goals such as the precision measurement of oscillation parameters, CP violation in the leptonic sector, searches for proton decay and detection of astrophysical neutrinos from a wide range of sources. 
The dominant detection channel in HK is the inverse beta decay (IBD) process, $\bar{\nu}_e + p \rightarrow e^+ + n $ which makes the detector particularly suitable for rise-time analysis. Although nearly $90\%$ of the detected events in HK arise from this IBD channel, the detector is also sensitive to other neutrino flavors through elastic neutrino--electron scattering, $\nu_X + e^- \rightarrow \nu_X + e^- $. Thus, due to the large detector mass, a significant number of electron neutrinos from the neutronization burst can be detected in HK.  



\textit{JUNO:}
The Jiangmen Underground Neutrino Observatory (JUNO), which recently started its operation, is a 20kt liquid scintillator detector located in Jiangmen, China~\cite{JUNO:2015zny}. 
The detector is planned to address various unknowns of the neutrino sector, including the mass hierarchy, precision measurement of oscillation parameters, detection of supernova neutrinos, atmospheric neutrinos, nucleon decay, etc. Similar to HK, the primary detection channel of JUNO is IBD, $\Bar{\nu_e} + p \rightarrow e^+ + n$ and suitable for the rise-time analysis. While we particularly focus on the IBD channel in JUNO, there are other subdominant interaction channels such as $\nu_e +  ^{12}C \rightarrow e^- + ^{12}N$,  $\Bar{\nu_e} +  ^{12}C \rightarrow e^+ + ^{12}B$, $ \nu + ^{12}C \rightarrow \nu + ^{12}C^\star$, and $\nu + e^- \rightarrow \nu + e^-$. However, the contribution of these channels is found to be negligible to the rise-time electron antineutrino detection.

To compute the neutrino events from a CCSN at the detectors discussed above, we use the public code ``SNOwGLoBES: SuperNova Observatories with GLoBES'' \cite{SNoWGLoBES}.  This code is designed to generate supernova neutrino events in different detectors like water Cherenkov, scintillator, liquid argon and lead. The code computes the interaction  and smeared rates for each interaction channel of a particular detector as function of energy. It uses the ``GLoBES'' library to calculate the event rates.

Note that SNOwGLoBES does not include neutrino oscillations. Therefore, we  perform the flavor conversion calculation outside SNOwGLoBES using the equations given in Sec.~\ref{subsec:neutrino_flux}, where the un-oscillated neutrino fluxes at source are obtained with Eq.~\ref{eq:source_flux}.   These oscillated neutrino fluxes are then fed as input to SNOwGLoBES. 
In the following, we discuss the SNOwGLoBES generated events for the detectors and analyze these results for possible implications on neutrino mass ordering.

\section{Results}
\label{sec:results}
Detection of neutrinos emitted during the neutronization burst and accretion phases from the next Galactic SN at the above mentioned detectors will help in identifying the true neutrino MO (NO and IO) as well as any imprints of the FE scenario. For probing these oscillation scenarios,  analysis of the plausible respective phenomenological signatures at the detectors becomes crucial. Thus, in the following, we compute the number of events at the three detectors DUNE, HK and JUNO for different oscillation scenarios as well as different progenitor models. Subsequently, we analyze the feasibility of discrimination between the three oscillation scenarios for different progenitor models.

\subsection{Neutronization Burst}
\label{subsec:nu_burst}
\begin{figure}
    \centering
    \includegraphics[width=\textwidth]{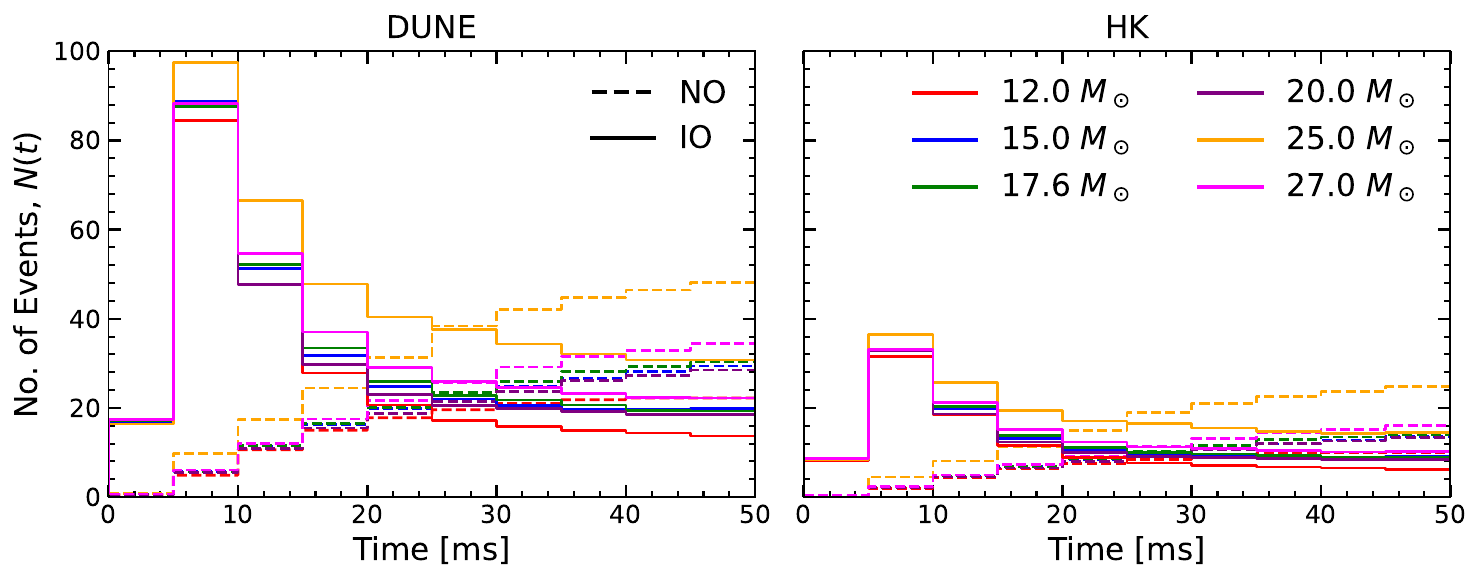}
    \caption{
    Expected number of neutronization burst $\nu_e$ events in DUNE (left) and HK (right) for different progenitor masses. The dashed and solid lines correspond to NO and IO, respectively.} 
    \label{fig:nu_burst}
\end{figure}

The flux of $\nu_{e}$ from a supernova can have different configurations for the two mass ordering scenarios (NO and IO) as shown in Eqs.~\ref{eq:NO_nue} and \ref{eq:IO_nue}. This mass ordering signature of the neutronization burst has remained one of the most prominent astrophysical neutrino phenomenology aimed by the upcoming (HK, DUNE) and running (JUNO) neutrino detectors.  HK, DUNE with their large volume would have excellent sensitivity to these early electron neutrino ($\nu_e$) emission phase, lasting about $20-30$ ms. Indeed, they would be able to distinguish the mass ordering scenarios. In the Fig.~\ref{fig:nu_burst}, we plot the expected number of $\nu_{e}$ events as a function of time for these two scenarios at DUNE (left) and at HK (right). The progenitor mass of the different Garching simulation models is mentioned in the figure legends. The distance of the Galactic supernova is taken to be $10$ kpc. Though the interaction channels in the two detectors are different, the time signature of the events at both detectors follows a similar pattern. The `$\nu_e$ burst' is clearly prominent for the IO as the IO flux (\ref{eq:IO_nue}) depends on the initial $\nu_e$ flux. In particular, the $\nu_{e}$ flux for IO is a mixture of the unoscillated $\nu_{e}$ ($ 31\% $) and $\nu_{x}$ ($69 \%$) flux, resulting in the peaked signature. For NO, the flux at the detector is being 
independent of the initial $\nu_e$ flux, the burst signature disappears in both  detectors. 
Among the two experiments, DUNE can detect more $\nu_{e}$ than HK. This is because the detection channel of DUNE involves charged current interaction with Argon. Whereas $\nu_{e}$ can be detected in HK by elastic scattering on electrons and, by charge and neutral current interaction with oxygen. These elastic scatterings on electrons dominate up to $20$ MeV and above it the interactions with oxygen dominate.  The supernova neutrino spectra peaks around $8-12$ MeV and then falls rapidly above $20$ MeV. Therefore, the dominant detection channel of $\nu_{e}$ from supernova at HK is elastic scattering with electrons. The cross-section for $\nu_{e}$ scattering on electrons is significantly smaller than that of charged current interaction in DUNE. Thus, the number of events at DUNE is expected to be larger than that of HK.
The progenitor mass dependence of the $\nu_e$ burst signal is also shown for both the detectors.   All the models show clear separation between NO and IO for both DUNE and HK. However, the event spectra of a particular mass ordering (NO or IO) for different progenitor masses are close to each other and the $\nu_e$ burst detection alone, might not be able to probe progenitor mass accurately.

\begin{figure}
    \centering
    \includegraphics[width=\textwidth]{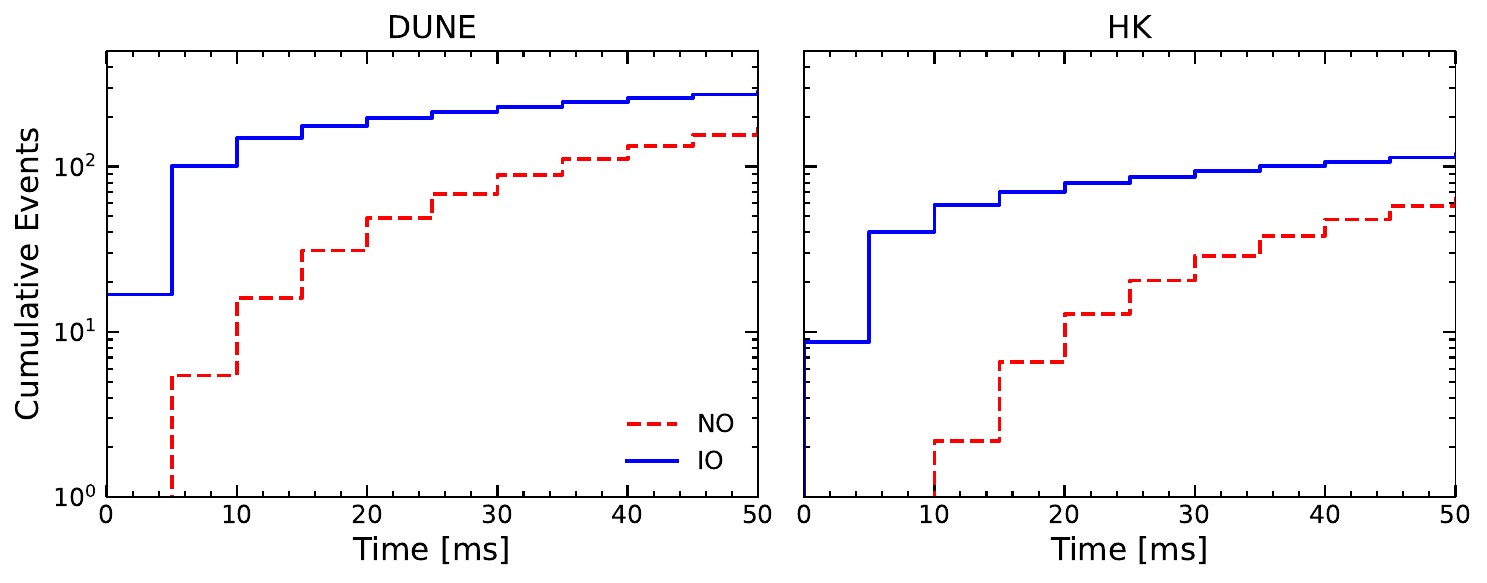}
    \caption{Expected number of  cumulative neutronization burst $\nu_e$ events in DUNE (left) and HK (right) for $12 M_{\odot}$.  The red dashed and blue solid lines correspond to NO and IO, respectively.} 
    \label{fig:nu_burst_cum}
\end{figure}

In addition to the comparison of the bin-by-bin event spectra, the spectra of bin-by-bin cumulative events can be another phenomenological possibility for distinguishing the true MO at the detectors. Therefore, we also show the spectra of bin-by-bin cumulative events in Fig.~\ref{fig:nu_burst_cum} for the $12~\rm M_{\odot}$ progenitor model at both DUNE (left) and HK (right). Although the sharp $\nu_e$ peak in the IO case disappears because of the accumulation of events over consecutive bins, the separation between the NO and IO spectra becomes larger than that observed in the bin-by-bin event spectra.

While Fig.~\ref{fig:nu_burst} and Fig.~\ref{fig:nu_burst_cum}   already show a clear distinction between the mass ordering scenarios, a statistical quantification of their separation is essential for a more robust assessment. In Sec.~\ref{sec:statistical_analysis}, we extend the neutronization burst signature detection with a detailed statistical analysis.


\begin{figure}
    \centering
    \includegraphics[width=\textwidth]{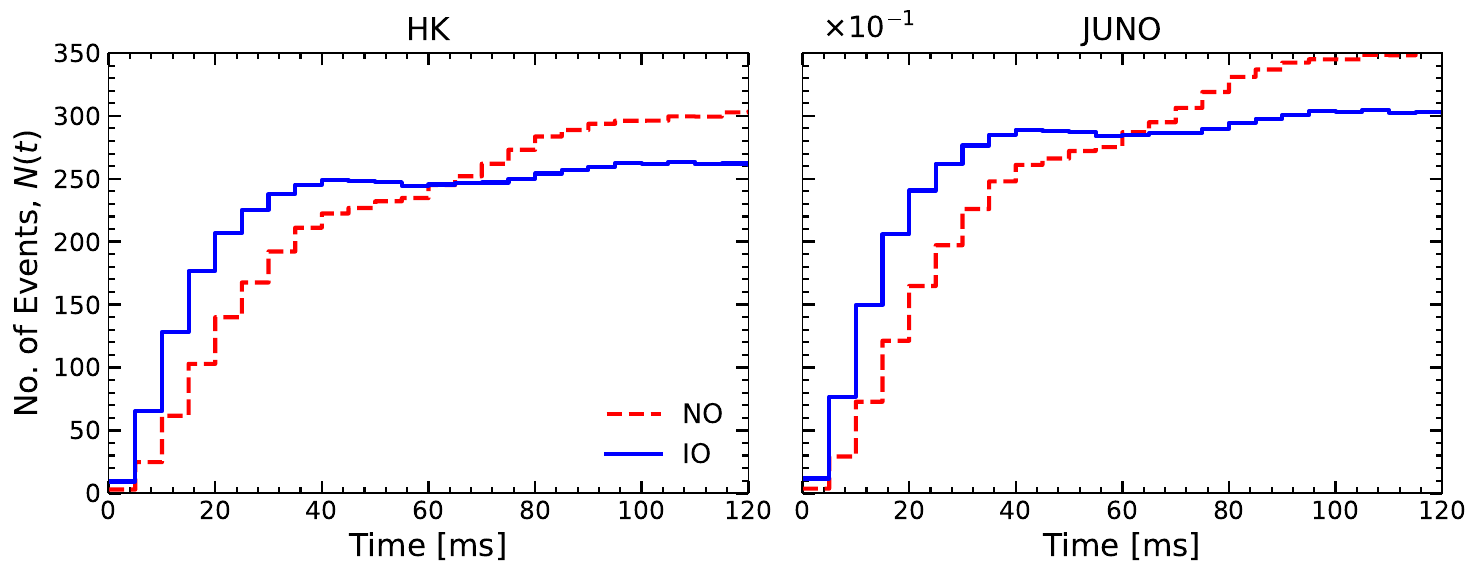}
    \caption{Expected number of $\bar{\nu}_e$  events from the accretion phase in HK (left) and JUNO (right) for a progenitor of $12~\rm M_{\odot}$. The red dashed and blue solid lines correspond to NO and IO, respectively.  }
    \label{fig:rise_time1}
\end{figure}

\begin{figure}
    \centering
    \includegraphics[width=\textwidth]{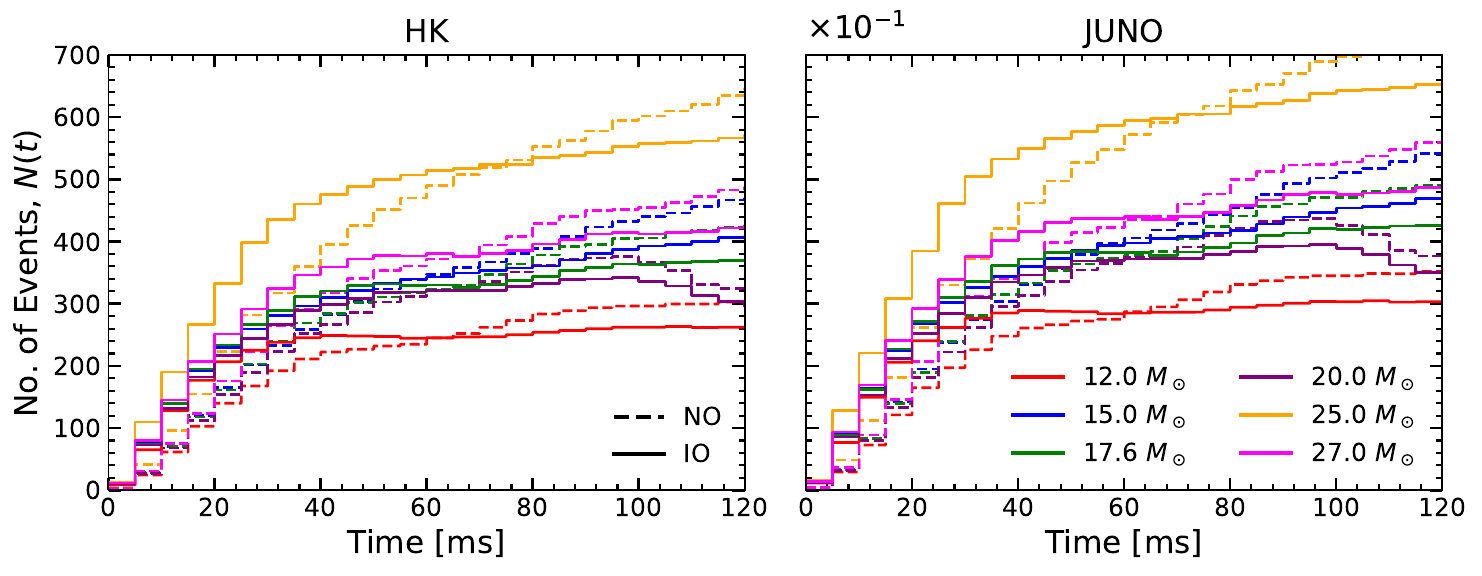}\\
    \caption{Expected number of $\bar{\nu}_e$ events  from the accretion phase in HK (left) and JUNO (right) for different mass ordering scenarios (NO: dashed line and IO: solid lines)  and different progenitor masses (shown by different colors).}
    \label{fig:rise_time2}
\end{figure}



\begin{figure}
    \centering
    \includegraphics[width=\textwidth]{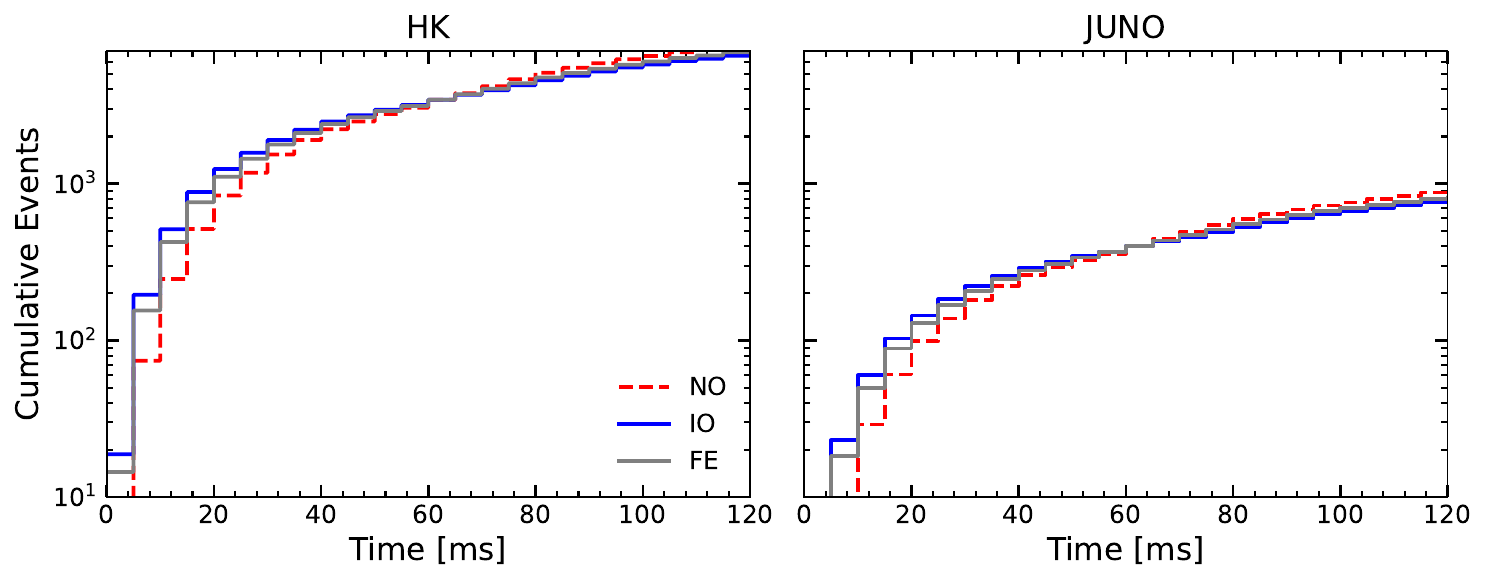}~
    \caption{Expected number of $\bar{\nu}_e$ cumulative events from the accretion phase in HK (left) and JUNO (right). The red dashed, blue solid, and gray solid lines correspond to the NO, IO, and FE, respectively.}
    \label{fig:Cumulative}
\end{figure}

\begin{figure}
    \centering
    \includegraphics[width=\textwidth]{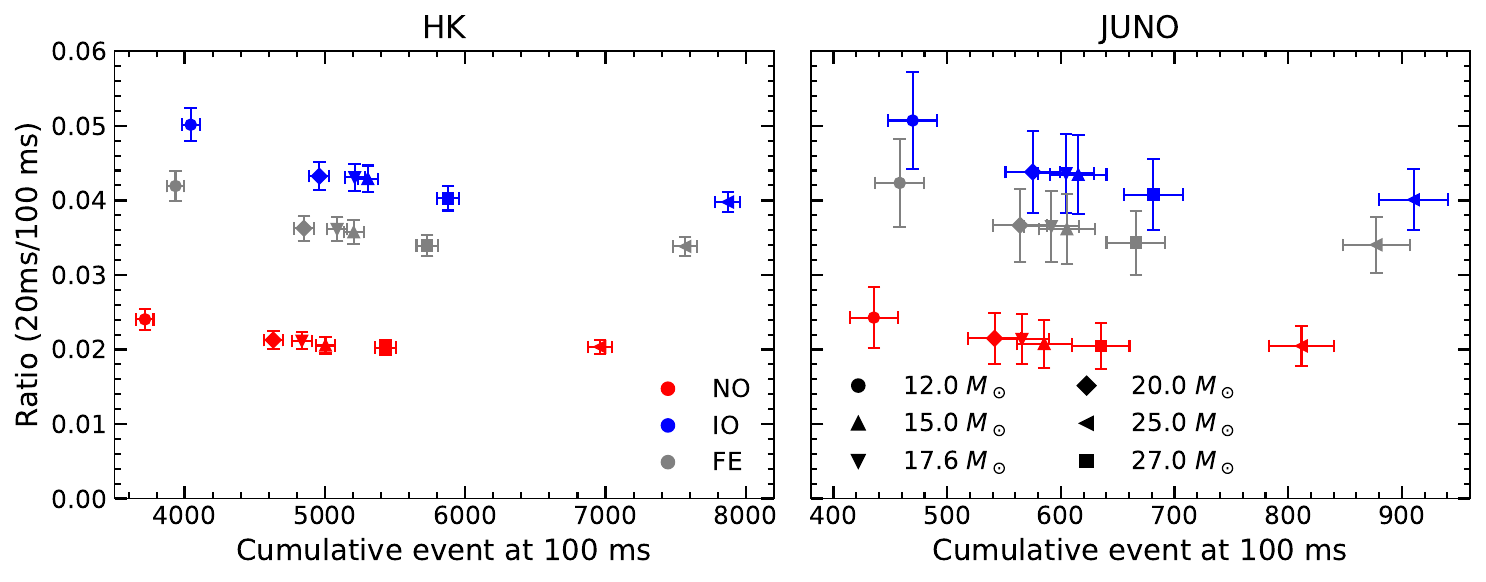}
    \caption{ Ratio of cumulative events at 20 ms to cumulative events 100ms vs cumulative events at 100ms for HK (left) and JUNO (right) for different progenitor masses. The NO, IO, and FE oscillation scenarios are represented by red, blue, and gray colors, respectively.
    }
    \label{fig:ratio_plot}
\end{figure}

\subsection{Rise time analysis}
\label{subsec:rise_time}

During the neutronization burst period, the production of neutrino species other than the $\nu_e$ are suppressed due to high electron degeneracy. When this degeneracy falls due to deleptonization, rapid production of $\bar{\nu}_{e}$ and $\nu_{x}$ begins. The luminosity of $\nu_{x}$ rises faster than that of $\bar{\nu}_{e}$ (see Sec.~\ref{sec:mass_ordering} for details). In fact, this feature of faster rise of the $\nu_{x}$ may result in unique time dependent signatures due to the flavor mixing.  The detectors sensitive to $\bar{\nu}_{e}$ and/or $\nu_{x}$ neutrino flavors have the possibility of detecting this rise time feature.  The IBD detectors, like IC, HK, JUNO with $\bar{\nu}_e$ being the main detection channel are best suited to detect such a time dependent signature. Since the SN neutrino flux at these IBD detectors has the signature of flavor oscillations, this indeed opens up the possibility of detecting the mass ordering in the early SN $\bar{\nu}_e$ flux.

For our oscillation scenario the $\bar{\nu}_{e}$ flux at the detectors (see Eqs.~\ref{eq:NO_nubare} and \ref{eq:IO_nubare} ) clearly have different characteristics. The MSW oscillated $\bar{\nu}_{e}$  flux at the detector for the IO, due to its pure $\bar{\nu}_{x}$ dependence, will follow the characteristics of the initial $\bar{\nu_{x}}$ flux. 
On the other hand, the $\bar{\nu}_{e}$  flux at the detector for the NO, will have the characteristics of both the initial $\bar{\nu}_e$ and the $\bar{\nu}_x$ flux. Thus at the IBD  detectors, like IC, JUNO and HK,  the oscillated $\bar{\nu}_{e}$ flux for the IO scenario grows faster than the NO case, and can reveal the rise time feature.
In the following, we compute the expected number of $\Bar{\nu}_{e}$ events for both JUNO and HK, as a function of time with a resolution of $5$ ms for different SN progenitor models assuming a source distance of $10$ kpc.


The Fig.~\ref{fig:rise_time1} panels show the expected $\Bar{\nu}_e$ events at HK (left) and JUNO (right) for the $12 M_{\odot}$ progenitor mass. The continuous and dashed lines correspond to NO and IO, respectively. The dashed dot lines correspond to the FE scenario. Clearly, the spectra of $\Bar{\nu}_e$ for IO case rises faster than that of the NO in both detectors for all models. 
HK will be capable of detecting a much larger number of events (by about an order) than JUNO, due to the larger size of HK than JUNO. This rise time feature of SN $\Bar{\nu}_{e}$, resulting in a clear separation between the different mass ordering cases at these detectors, is considered to be a crucial characteristic for the SN mass ordering studies \cite{Serpico:2011ir}. Note, the crossing between NO and IO spectra at around $60$ ms, it happens due to the fact that the production of $\Bar{\nu}_{e}$ becomes more efficient with time.  This is because the processes like positron capture and electron-positron annihilation start to contribute significantly, increasing the flux of $\Bar{\nu}_{e}$. 
However, this clean signature of the rise time may not remain true once we consider the different progenitor models. Indeed, the separation between the NO and IO cases is distinct for any particular progenitor mass model, but the separation for two different mass models can bring in degeneracy. This has been shown in the Fig.~\ref{fig:rise_time2} for both HK (left panel) and JUNO (right panel). The rise time feature for both the mass ordering has been shown for the six different progenitor models. The rise time cannot visibly distinguish the mass ordering cases for different progenitor models, unlike the neutronization phase. For example, the NO case of $15~\rm M_{\odot}$ model is difficult to distinguish from the IO case of $17.3~\rm M_{\odot}$ model.
Further statistical analysis and different index proposals 
were made to untangle this mass model dependent `degeneracy' while detecting mass ordering in the rise time feature.  

One proposal to understand this distinguishability of the different oscillation scenarios among different progenitor models is to utilize the rise time feature of the cumulative events. In comparison to the rise time feature of the events, the time dependent cumulative events will have much larger number of events in each bin. For example consider the Fig.\ref{fig:Cumulative} for the  $12~\rm M_{\odot}$ case, the cumulative events have been plotted for both HK (left panel) and JUNO (right panel). This is expected to result in smaller statistical error for each bin in comparison to the simple rise proposal (see Fig. \ref{fig:rise_time1}). The cumulative events for the IO grow much faster than the NO case.
We have also plotted the FE scenario for both the detectors. Though the separation is better than the simple event rise time, solving the progenitor mass model degeneracy problem will depend on the detector characteristics and size. In fact, the possibility of the FE scenario complicates things further than analyzed earlier in a simple two mass ordering case \cite{Serpico:2011ir}.

The proposal of using a cumulative index, based on the neutrino emission characteristic time scales, is considered to be useful to resolve this problem. 
The proposal is based on the realization that for almost all the different progenitor masses of the hydrodynamic SN simulations there are some common characteristic timescales. For example, the luminosity of $\nu_{x}$ rises faster than $\Bar{\nu}_{e}$ and reaches a maximum at around $20$ ms and then the luminosity of both these flavors saturates around $100$ ms (see, Fig. \ref{fig:light_curves}). As discussed in Sec.~\ref{sec:SN_models}, the faster rise of $\nu_{x}$ luminosity within $\sim 20$ ms is due to the high electron degeneracy in the SN core, leading to slow increase  of $\Bar{\nu}_{e}$ luminosity. However, the time  $100$ ms corresponds to the end of the free propagation of the shock. These physical properties are believed to be prevalent in SN models, however, the corresponding timescales may vary slightly depending on the models.
These characteristic time scales are found to be useful for probing the true mass ordering signals. The proposed index is still a measure of the growth rate of the fluxes in the different oscillation scenarios, and measures the ratio between the cumulative events at the $\nu_{x}$ maxima and saturation time scales.
\begin{equation}
    {\rm Ratio}  = \frac{{\text{Cumulative events at 20 ms}} }{{\text{Cumulative events at 100 ms}}}\,.
\label{eq:ratio}
\end{equation}

In Fig.~\ref{fig:ratio_plot} we have plotted this ratio as a function of the cumulative events at $100$ ms for the different progenitor models.  The left panel is for HK and the right panel is for JUNO.  Each of these points in the figure represents this ratio for a specific oscillation scenario. The color coding remains the same as the previous plots, NO cases in red, IO ones in blue and the FE scenario plotted in gray. The statistical errors for each of the points are also shown in the plot, where we consider the propagated errors from the $1~\sigma$ Poisson uncertainties in the events. In comparison to JUNO, HK will have much better sensitivity, as the statistical error in HK is much lower due to its larger fiducial volume and larger expected events per time bin. 
Evidently, for both detectors, the cumulative nature of the quantities help in the error estimates in comparison to the simple bin-by-bin event analysis.   
The mass ordering scenarios, NO, IO and FE are well separable in HK for a specific progenitor mass model and also between different mass models. However, the discrimination is poor in JUNO due to the larger uncertainties, in particular the FE case may not have a clear separation from the IO scenario. This feature that the difference between FE and IO is smaller than the difference between FE and NO, is also there in HK. However, this situation becomes worse for JUNO as the error bars for FE and IO overlap. Nevertheless, JUNO might still be able to see a clear separation between the FE/IO and NO.  

Our results based on the  time variation analysis for a future Galactic supernova neutrino events show optimistic possibility of probing the neutrino mass ordering in the DUNE, HK and JUNO detector. Features like $\nu_e$ burst and rise of $\bar{\nu}_{e}$ at these detectors provide us with different insights into the mass ordering problem. All the three detectors are found to have good sensitivity for the neutrino  mass ordering in their respective detection channels. This difference in the detection channels also raises the issue of probing the statistical significance in further detail. The SN progenitor mass model is one crucial source of uncertainty. In particular, the neutronization burst phase in comparison to the accretion phase's rise time feature is found to be least affected by this SN mass uncertainty. Our preliminary results with the proposal of the cumulative event ratio in rise time show encouraging promise. However, a further statistical analysis based on the probability of identification between the same and different mass models is required to settle the model the degeneracy.   In the following, we describe the statistical method to estimate this probability of identification for all three detectors.   


\begin{figure}
    \centering    
    \includegraphics[width=0.49\linewidth]{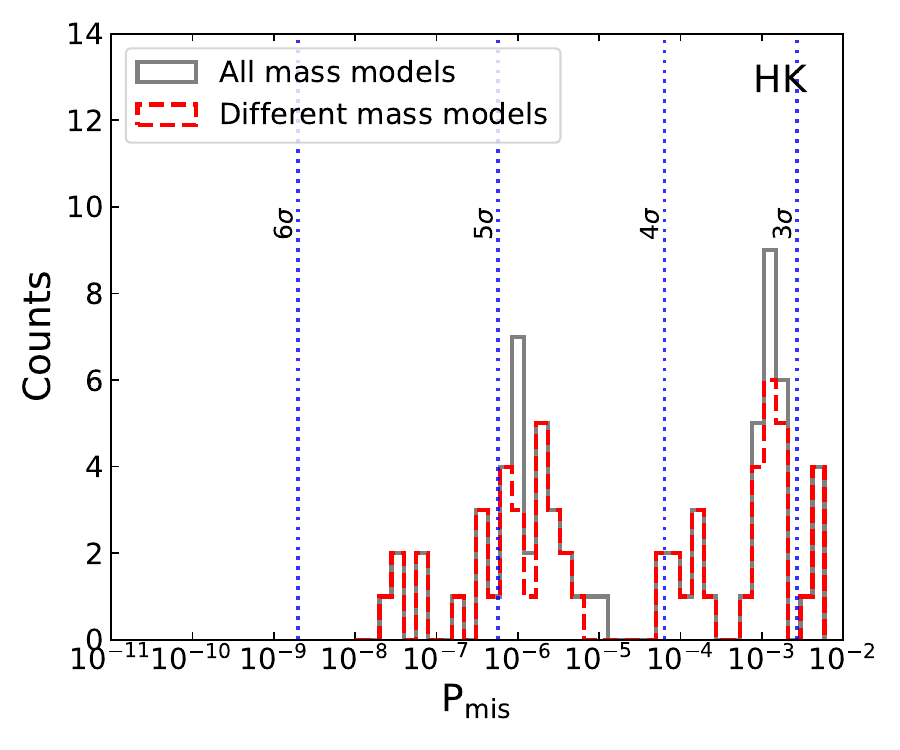}
    \includegraphics[width=0.49\linewidth]{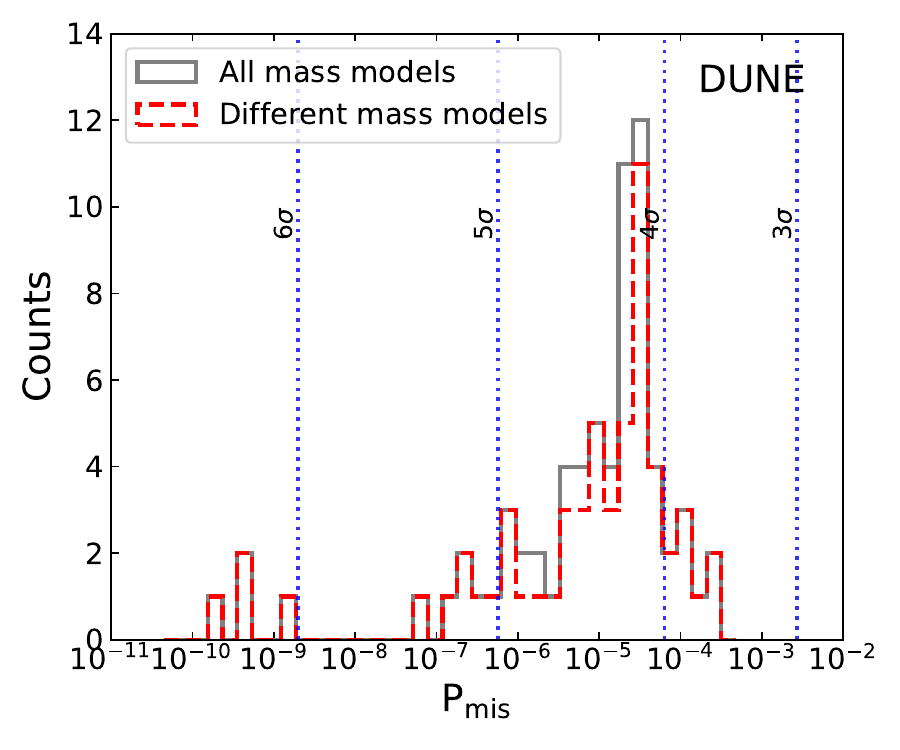}
    \includegraphics[width=0.49\linewidth]{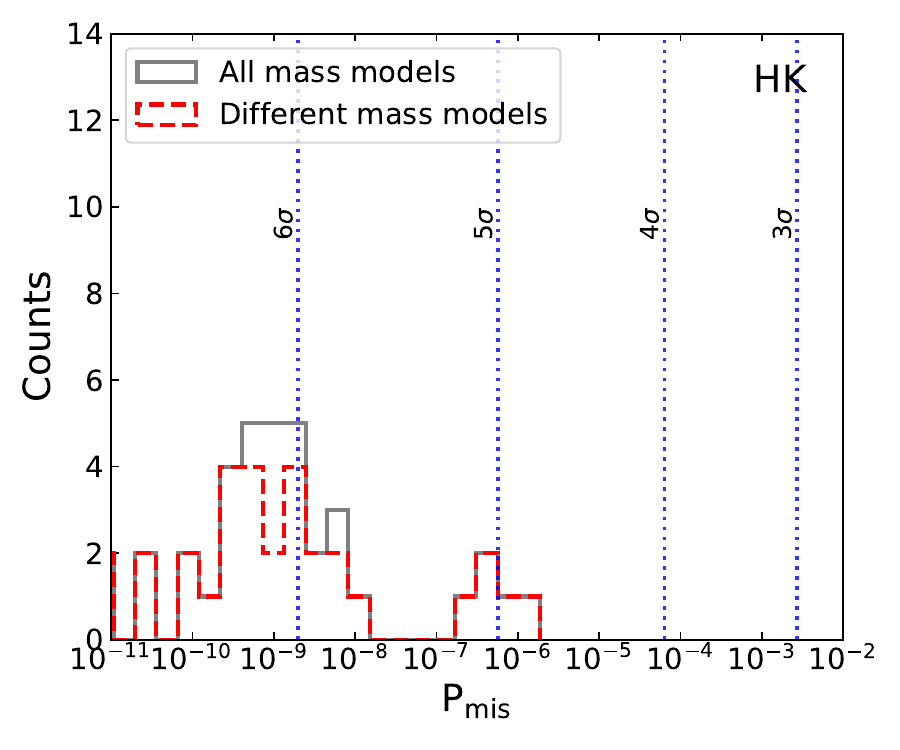}
    \includegraphics[width=0.49\linewidth]{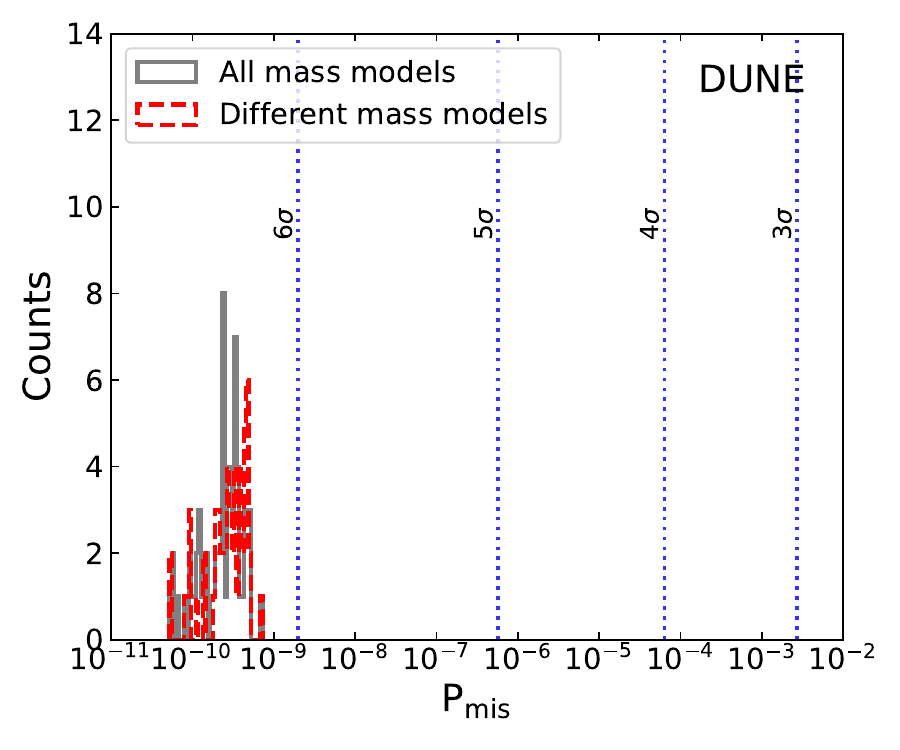}    
    \caption{MO misidentification probabilities ($\rm P_{mis}$) for the neutronization burst in HK (left) and DUNE (right). The gray and red bins correspond to All mass models and Different mass models, respectively. The blue vertical lines show the different confidence levels ($3 \sigma$, $4 \sigma$, $4 \sigma$, and $6 \sigma$). The top and bottom panels depict the $\rm P_{mis}$ for bin-by-bin events and bin-by-bin cumulative events approaches, respectively.}
    \label{fig:CL-Nuburst}
\end{figure}

\begin{table}[h]
\centering
\begin{tabular}{llcccc}
\hline
Detectors & Model &  $2\sigma$ & $3\sigma$ & $4\sigma$ & $5\sigma$ \\
\hline
\multirow{2}{*}{HK}
 & All mass models        &  100 & 93 & 53 & 14 \\
 & Different mass models &  100 & 92 & 52 & 17 \\
\hline
\multirow{2}{*}{DUNE}
 & All mass models        & 100 & 100 & 88 & 15 \\
 & Different mass models & 100 & 100 & 85 & 18 \\
\hline
\end{tabular}
\caption{Percentage of models surviving different statistical significance thresholds for HK and DUNE for the neutronization burst using bin-by-bin event analysis.}
\label{tab:hk_dune_sigma}
\end{table}

\begin{table}[h]
\centering
\begin{tabular}{llccccc}
\hline
Detectors & Model &  $2\sigma$ & $3\sigma$ & $4\sigma$ & $5\sigma$ & $6\sigma$ \\
\hline
\multirow{2}{*}{HK}
 & All mass models        & 100 & 100 & 100 & 97 & 83 \\
 & Different mass models & 100 & 100 & 100 & 97 & 83 \\
\hline
\multirow{2}{*}{DUNE}
 & All mass models        & 100 & 100 & 100 & 100 & 100 \\
 & Different mass models & 100 & 100 & 100 & 100 & 100 \\
\hline
\end{tabular}
\caption{Percentage of models surviving different statistical significance thresholds for HK and DUNE for the neutronization burst bin-by-bin event cumulative analysis.}
\label{tab:hk_dune_sigma_cum}
\end{table}

\begin{figure}[!h]
    \centering
    \includegraphics[width=0.49\textwidth]{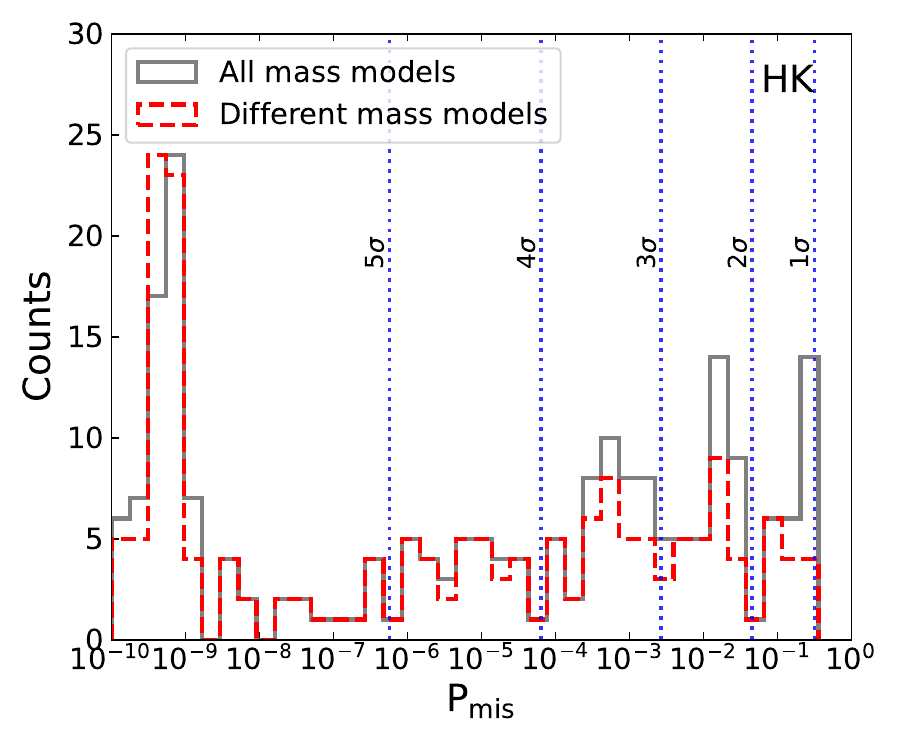}
    \includegraphics[width=0.49\textwidth]{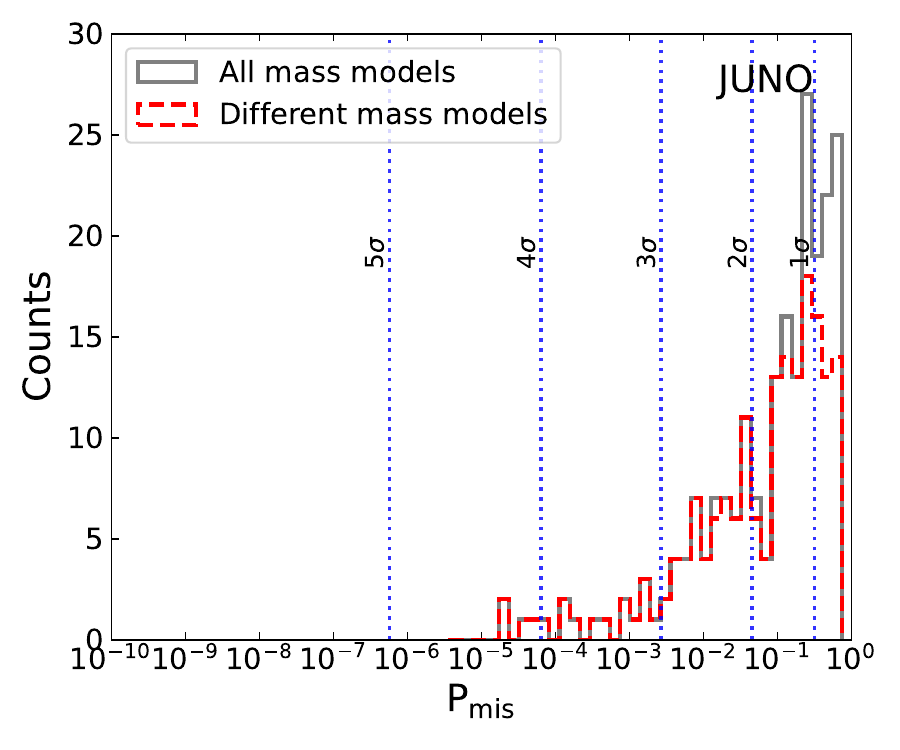}
    \includegraphics[width=0.49\textwidth]{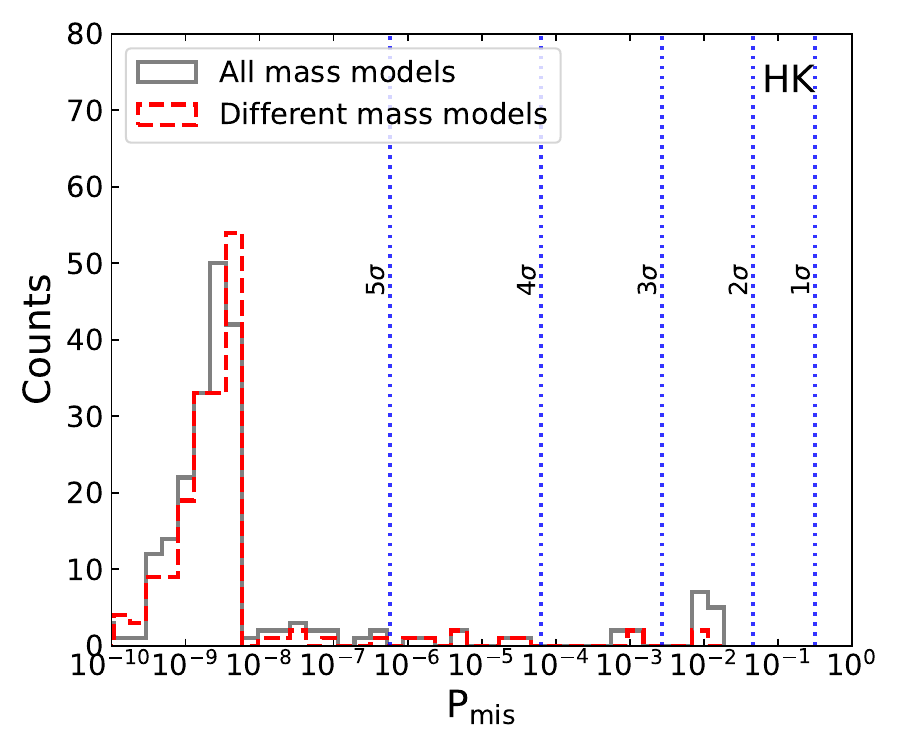}
    \includegraphics[width=0.49\textwidth]{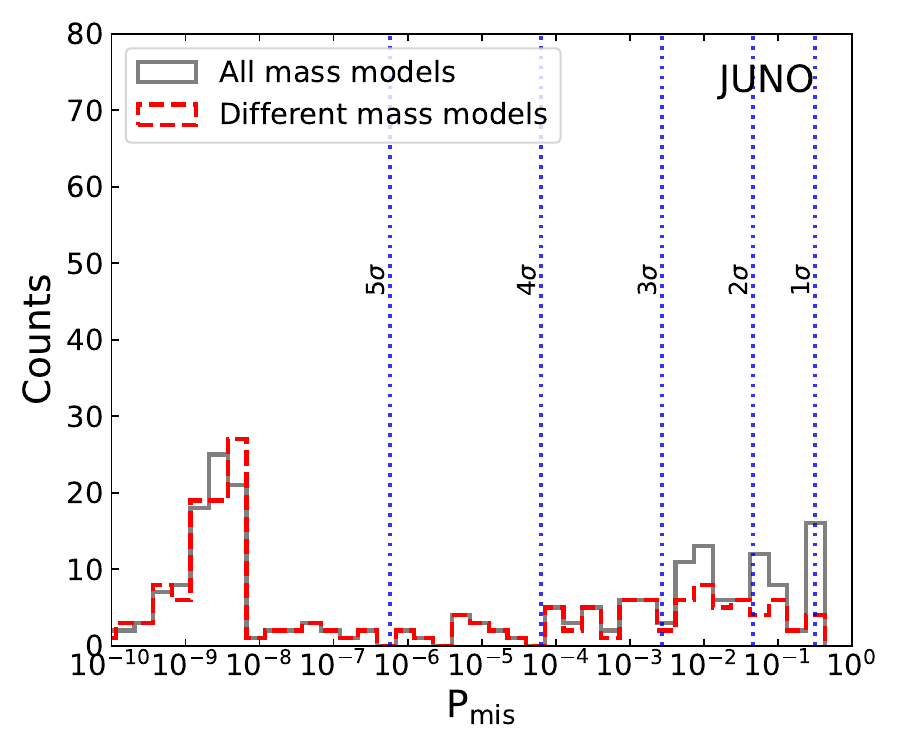}
    \includegraphics[width=0.49\textwidth]{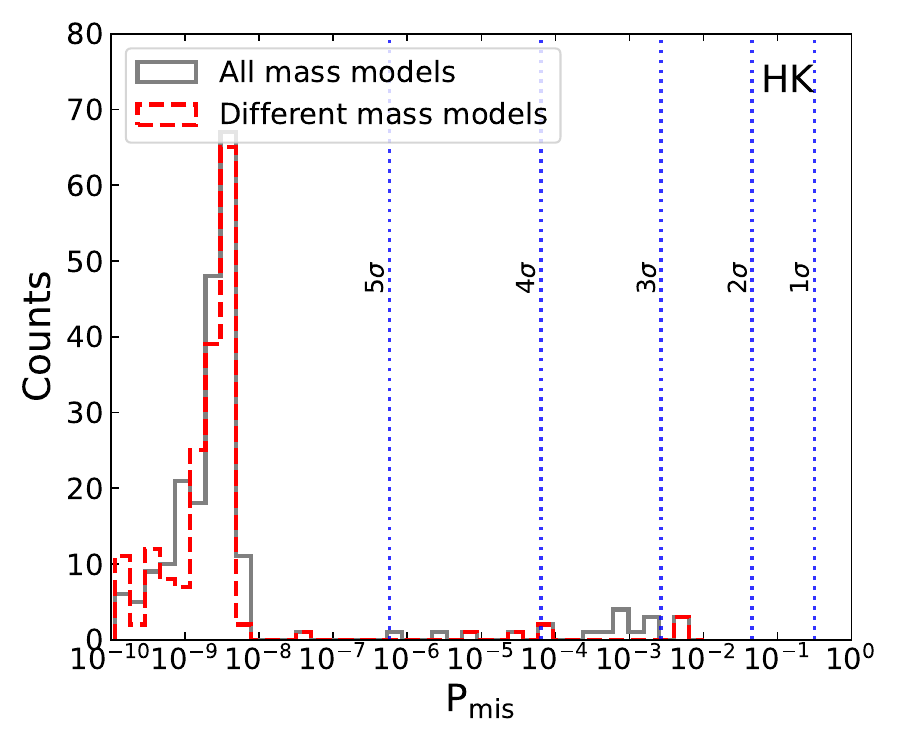}
    \includegraphics[width=0.49\textwidth]{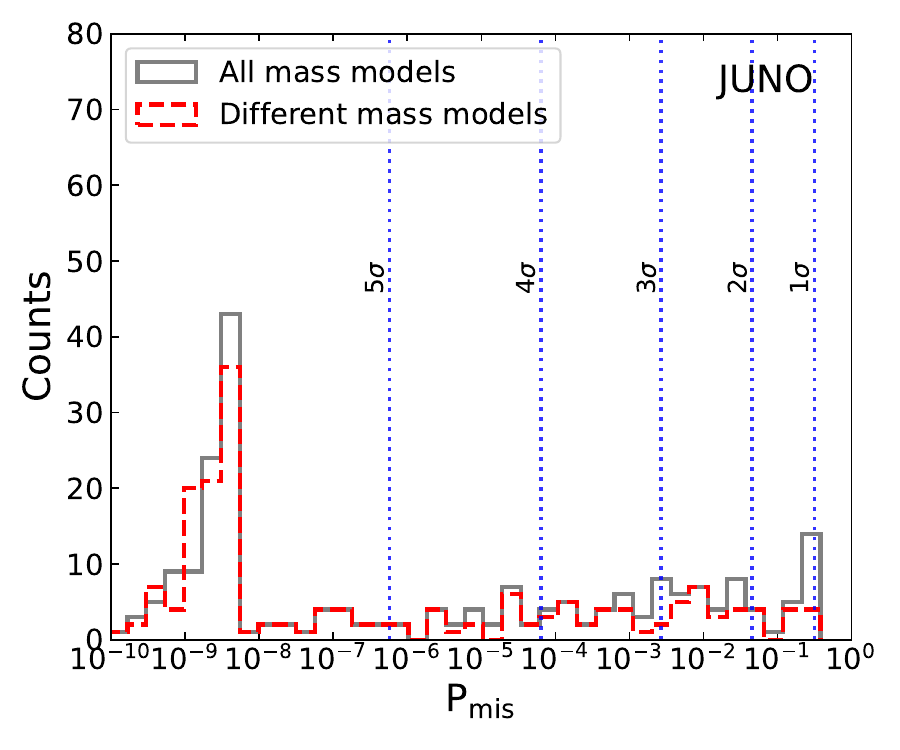}
    \caption{MO misidentification probabilities ($\rm P_{mis}$) for the neutronization burst in HK (left) and JUNO (right). The gray and red bins correspond to All mass models and Different mass models, respectively. The blue vertical lines show the different confidence levels ($(1-5) \sigma$). \textit{Top panel:} For bin-by-bin event analysis. \textit{Middle panel:} For by bin cumulative event analysis. \textit{Bottom panel:} For ratio of cumulative events at two different instants ($20$ ms and $100$ ms). }
    \label{fig:sigma_rise_time}
\end{figure}

\begin{table}[h]
\centering
\begin{tabular}{llccccc}
\hline
Detector & Models & $1\sigma$ & $2\sigma$ & $3\sigma$ & $4\sigma$ & $5\sigma$ \\
\hline
\multirow{2}{*}{HK}
 & All mass models        & 98 & 88 & 72 & 51 & 37 \\
 & Different mass models & 99 & 92 & 78 & 60 & 44 \\
\hline
\multirow{2}{*}{JUNO}
 & All mass models        & 73 & 33 & 8 & 2 & 0 \\
 & Different mass models & 78 & 39 & 10 & 3 & 0 \\
\hline
\end{tabular}
\caption{Percentage of models surviving different statistical significance thresholds for HK and JUNO for bin-by-bin event analysis for rise time.}
\label{tab:hk_juno_sigma}
\end{table}

\begin{table}[h]
\centering
\begin{tabular}{llccccc}
\hline
Detector & Models & $1\sigma$ & $2\sigma$ & $3\sigma$ & $4\sigma$ & $5\sigma$ \\
\hline
\multirow{2}{*}{HK}
 & All mass models        & 100 & 100 & 94 & 93 & 90 \\
 & Different mass models & 100 & 100 & 99 & 98 & 94 \\
\hline
\multirow{2}{*}{JUNO}
 & All mass models        & 94 & 82 & 64 & 52 & 46 \\
 & Different mass models & 99 & 91 & 76 & 62 & 55 \\
\hline
\end{tabular}
\caption{Percentage of models surviving different statistical significance thresholds for HK and JUNO for bin-by-bin cumulative event analysis for rise time.}
\label{tab:hk_juno_sigma_cum}
\end{table}

\begin{table}[h]
\centering
\begin{tabular}{llccccc}
\hline
Detector & Models & $1\sigma$ & $2\sigma$ & $3\sigma$ & $4\sigma$ & $5\sigma$ \\
\hline
\multirow{2}{*}{HK}
 & All mass models        & 100 & 100 & 98 & 93 & 91 \\
 & Different mass models & 100 & 100 & 98 & 97 & 96 \\
\hline
\multirow{2}{*}{JUNO}
 & All mass models        & 99 & 90 & 75 & 63 & 52 \\
 & Different mass models & 98 & 94 & 82 & 71 & 62 \\
\hline
\end{tabular}
\caption{Percentage of models surviving different statistical significance thresholds for HK and JUNO for the ratio index of rise time.}
\label{tab:hk_juno_sigma_ratio}
\end{table}

\section{Statistical analysis}
\label{sec:statistical_analysis}

For the statistical significance of the mass ordering identification between different progenitor masses we begin with a chi square analysis. From these chi-square distributions we estimate the probability to identify specific oscillation scenario both within the progenitor model and among the different models. 
Thus for this purpose to distinguish the two mass ordering scenarios as well as different progenitor models, the chi square difference is defined in the following manner,

\begin{equation}
    \chi^2=2 \sum_{i=1}^{N_{bin}} \left[N_{\rm i }^{Test} -N_{i}^{True}\left\{1+\log\left( \frac{N_{\rm i}^{Test}}{N_{i}^{True}} \right)\right\} \right] \,
    \label{eq:chisq}
\end{equation}
where, $i$ represents the number of time bins. The number of bins are different for the neutronization burst and the rise time analysis. In the case of neutronization burst for the expected $5$ ms time bins at DUNE and HK, $i=1-6$ corresponding to the typical burst time scale of $30$ms. For the rise time the flux saturation time scales in all the progenitor models are about $100$ ms, corresponding to $i=1-20$. $N_{\rm i}^{True}$ represents events in the $i^{\rm th}$ bin for the true model with a particular choice of the progenitor mass model ($\it mod$) and mass ordering oscillation scenario ($\it MO$). This true model is tested with respect to all the other possible combinations of $mod$ and $MO$, the $i^{\rm th}$ bin events for these test scenarios are  represented by the $N_{\rm i}^{Test}$. The degrees of freedom (DOF) for these chi square distributions are $N_{bin}-1$ and the reduced chi square is defined as $\chi^2_{red} = \chi_{\rm O,E}^2/\mathrm{DOF}$.

The expected events for each choice of true case ($mod$, $MO$) are generated with the SNOwGLoBES, based on the inputs from the hydrodynamic simulation data. The test events are obtained by randomizing the relevant SNOwGLoBES templates for all the possible combinations of choices ($mod$, $MO$). We consider these randomized events for each of these test cases to follow Poisson statistics and are generated using $1 \sigma$ Poisson uncertainties, i.e, $\sqrt{N_i}$ in the interval $[(N_i -\sqrt{N_i}), (N_i+\sqrt{N_i})]$. Finally, using the $\chi^2$ values for different model combinations, we compute the respective $p$ values.  Given the diversity and further possibility of different SN models (mod) it is important to probe the possibility of misidentification of MO with 
each of these different model combinations. Indeed, the $p$ values represent these probabilities of the mass ordering misidentification, i.e., measure of the possibility to confuse the distinct MO identification signal with the uncertainty due to SN models. Using these $p$ values, the corresponding confidence levels ($\sigma$) can be determined assuming Gaussian statistics. 

The statistical analysis of the MO identification with the neutronization burst signal is accessible through only two the oscillation scenarios, i.e., IO and MO. This is due to the fact that the large lepton number asymmetry in this early phase of neutronization burst can suppress any collective flavor oscillation. Hence, we avoid the fast collective oscillations derived FE scenario in this phase \cite{Capozzi:2018rzl}. Thus, for each of the six progenitor mass models there are two oscillation possibilities, i.e., there can be $12$ `test' cases. Against each of these test cases there can be $6$ true cases excluding the same MO scenarios, making a total counts of $72$  model combinations. In the top panel of Fig.~\ref{fig:CL-Nuburst}, we show the model counts versus the MO misidentification probability for the bin-by-bin events spectra depicted in Fig.~\ref{fig:nu_burst}. The left and right panels correspond to the detectors HK and DUNE, respectively. The vertical blue dotted lines depict different confidence levels, i.e., (3-6) $\sigma$. The gray lines show the MO misidentification probability for all the possible \textit{mod} combinations ($72$), whereas the red lines depict the misidentification probabilities for \textit{mod} cases excluding the combinations of same progenitor mass ($60$). For both scenarios the MO misidentification probabilities are extremely small. Clearly, DUNE yields smaller misidentification probabilities compared to HK due to the DUNE's better sensitivity to electron neutrinos. In particular, DUNE shows misidentification probabilities $\leq 5 \times 10^{-4}$  for all the \textit{mod}-MO combinations. This implies that the true MO can be identified in DUNE with confidence level $\gtrsim 3 \sigma$. In fact, the confidence level is  $ \sigma \sigma$ for $\sim 88\%$ of the model counts. Whereas, the confidence level for HK is found to be  $\gtrsim 2.7 \sigma$ corresponding to misidentification probability $\leq 6 \times 10^{-3}$. For better visualization, we also show the percentage of models corresponding to different CL in Table~\ref{tab:hk_dune_sigma}.
While the statistical analysis based on the bin-by-bin event spectra already shows a robust distinction between the two MOs, the analysis using bin-by-bin cumulative events may further enhance the confidence level of MO discrimination. Therefore, we also present the MO misidentification probabilities obtained from the bin-by-bin cumulative events in the bottom panels of Fig.~\ref{fig:CL-Nuburst} for HK (left) and DUNE (right). The fractions of models corresponding to different confidence levels obtained from this method are reported in Table~\ref{tab:hk_dune_sigma_cum}.
Clearly, this method results in extremely small misidentification probabilities for both HK and DUNE. For all the models considered, DUNE (HK) can identify the true MO with a confidence level of $6~\sigma$ ($4~\sigma$).
Thus, the proposal that the `standard neutrino candle' signal of neutronization burst is a robust probe MO identification stands our statistical test. Indeed, the statistical significance consistent with the original neutronization burst-MO sensitivity analysis  \cite{Kachelriess:2004ds} with substantial improvement due to the improved clarity of the oscillation scenario and understanding of the detector sensitivities.

Given the usefulness of this statistical analysis method in estimating the true MO scenario in the neutronization burst phase,  we now also employ it for the rise time burst phase. However, unlike the neutronization  burst phase, the third flavor conversion possibility (FE) due to collective oscillations also comes into the picture in addition to the standard NO and IO scenarios. As discussed in Sec.~\ref{sec:SN_models}, this FE scenario behaves as an intermediate possibility between the NO and IO scenarios, making the true MO identification complicated than that of the neutronization burst phase. Nevertheless, we employ the same statistical analysis of misidentification probability of model discrimination  for the rise time, considering all three oscillation possibilities (NO, FE, and IO) for the detectors, HK and JUNO. Now, for this the total number of test cases becomes $18$ using the same counting method as before. Against each of these $18$  test cases, there can be $12$ true cases excluding the same MO scenarios. Thus, the total model counts for rise time is $216$ (All mass models). After subtracting the cases where two models share the same progenitor mass, the total counts reduces to $180$ (Different mass models). For both these model counting scenarios, we compute the misidentification probabilities and depict them in the top panel of Fig.~\ref{fig:sigma_rise_time}. The left panel (right) shows the misidentification probabilities for HK (JUNO).  
For the quality assessment  of these results, we also show different statistical significance ($n \sigma$, where $n=1-5$) in both plots by the vertical blue dotted lines. In addition, we compute the percentage of model counts for these different statistical significance and tabulated them in Table~\ref{tab:hk_juno_sigma}. 
Clearly, the misidentification probabilities are much larger i.e., statistically less significant than that obtained in the neutronization burst phase. This is due to the fact that the for the different $mod$s and MO combinations, it is possible to have small summed over bin-by-bin flux difference, even though the slopes are very different, see Fig.~\ref{fig:rise_time2}. Having included the FE scenario also reduces the statistical significance. Thus, this bin-by-bin event statistical analysis for true MO identification is not suitable for the rise time phase. 

The misidentification probabilities can be improved by increasing the separation between the models. This can be achieved by computing the bin-by-bin cumulative  events for each of the models as shown in Fig.~\ref{fig:Cumulative}.  Now, we compute the $\chi^2$ differences for all the model combinations ($216$ and $180$) discussed above and the corresponding misidentification probabilities using the same method. The misidentification probabilities are shown in the middle panel of Fig.~\ref{fig:sigma_rise_time} for both HK (left) and JUNO (right). Similar to the top panel of Fig.~\ref{fig:sigma_rise_time}, the different statistical significances have also been shown in the plots. Evidently, the comparison of the models through bin-by-bin cumulative events have resulted in smaller probabilities of misidentification compared to the bin-by-bin event analysis. For instance, true MO identification in HK (left plot) for all the model counts can be made at confidence level larger than $2 \sigma$ using this cumulative event approach. However, as expected the MO distinction in JUNO is found to be poorer compared to HK. For detailed comparison, the percentage of models corresponding to different confidence levels for both HK and JUNO have been listed in Table~\ref{tab:hk_juno_sigma_cum}.   Interestingly, $99\%$ ($94\%$) of the different mass models (all mass models) can be distinguished at confidence level $\geq 3 \sigma$ in HK. Despite poor statistical significance in JUNO, $76\%$ ($64\%$) of the different mass models (all mass models) are found to have confidence level $\geq 3 \sigma$. Thus, this method of bin-by-bin cumulative event analysis of the neutrino spectra during the rise time phase will be a crucial tool of probing the true MO.

As discussed in Sec.~\ref{sec:results}, the index defined by the ratio of cumulative events at two different instants ($20$ ms and $100$ ms) can serve as an effective indicator of the MO. This ratio index exhibits a clear separation between the standard NO and IO scenarios for both HK and JUNO, as shown in Fig.~\ref{fig:ratio_plot}. However, the distinction between the IO and IO scenarios is less pronounced compared to that between NO and IO.
To quantify the separation between different MO scenarios, we perform a chi-square analysis for this ratio index. Nevertheless, the chi-square definition given in Eq.~\ref{eq:chisq} is not applicable in this case, as the cumulative events at $20$ ms ($E_{20}$) and $100$ ms ($E_{100}$), used to construct the ratio, are correlated. Therefore, we adopt the following chi-square definition for correlated variables~\cite{Cowan:1998ji,Cowan:2018lhq},

\begin{equation}
\chi^2 =
\frac{1}{1-\rho^2}
\left[
\frac{\left(E_{20}^{\rm Test} - E_{20}^{\rm True}\right)^2}{\sigma_{20}^2}
+
\frac{\left(E_{100}^{\rm Test} - E_{100}^{\rm True}\right)^2}{\sigma_{100}^2}
-
\frac{2\rho \left(E_{20}^{\rm Test} - E_{20}^{\rm True}\right)
\left(E_{100}^{\rm Test} - E_{100}^{\rm True}\right)}
{\sigma_{20}\sigma_{100}}
\right] \ ,
\end{equation}

where, the correlation coefficient ($\rho$) is estimated in terms of $\mathrm{Cov}(E_{20}, E_{100})$, i.e, the covariance between $E_{20}$ and $E_{100}$, and given by  $\rho = \frac{\mathrm{Cov}(E_{20}, E_{100})}{\sigma_{20}\sigma_{100}}$.
 The test cases $E_{20}^{\rm Test}$ and $E_{100}^{\rm Test}$ are obtained by  the mean of $10^4$  randomly generated samples with corresponding standard deviations $\sigma_{20}$ and $\sigma_{100}$. 
 
Using this chi square definition, we compute the misidentification probabilities ($\rm P_{mis}$) for both model count scenarios (All mass models and Different mass models) and show the results in the bottom panel of Fig.~\ref{fig:sigma_rise_time} for both HK (left) and JUNO (right). In addition, the percentages of model counts corresponding to different confidence levels are shown in Table~\ref{tab:hk_juno_sigma_ratio}. Interestingly, the results of this statistical test are found to be significantly better compared to the bin-by-bin cumulative event case.  This can be clearly realized by comparing Table~\ref{tab:hk_juno_sigma_cum} and Table~\ref{tab:hk_juno_sigma_ratio}. For instance, in the case of all mass models using bin-by-bin cumulative events in HK (JUNO), only $94\%$ ($64\%$) of the models reach the $3\sigma$ confidence level, whereas this fraction improves to $98\%$ ($75\%$) when using the ratio index. Thus, the ratio index emerges as the most sensitive probe of the MO scenarios for the accretion phase, while the bin-by-bin cumulative events offer complementary insights. Note that the correlation coefficient, $\rho$, is negligible across all models; hence, assuming $\rho = 0$ yields similar results.

The  weaker sensitivity of the rise time observable to MO discrimination in comparison to the neutronization burst is due to the inclusion of the FE scenario. Interestingly, in the absence of the FE scenario, the rise-time sensitivity improves significantly, comparable to that of the neutronization burst~ $\sim 5~\sigma$. Nevertheless, our work emphasizes the impact of uncertainties arising due to self-induced collective oscillations during the accretion phase such as FE, which were not taken into account in earlier studies~\cite{Serpico:2011ir}. In addition, unlike previous analyses, our statistical framework enables us to quantify the impact of progenitor model uncertainties on MO discrimination.

In the above discussion, we have presented an elaborate statistical analysis of MO identification for both the neutronization burst and rise time. Overall, the neutronization burst being the standard SN candle is found to be the best tool for probing the true MO from the next Galactic SN.  While the rise time feature could also provide highly statistically significant information on the MO, and thus complements the results of the neutronization burst.


\section{Conclusion}
\label{sec:conclusion}
A future Galactic CCSN will provide an unprecedented
opportunity to probe fundamental properties of neutrinos, in particular the
neutrino MO. In this work, we have performed a comprehensive
analysis of the MO sensitivity by exploiting the early-time neutrino signal
from different phases of the SN evolution, at the DUNE,
HK, and JUNO detectors.

We have focused on two key observables arising from distinct emission phases:
the electron neutrino ($\nu_e$) neutronization burst and the rise-time behavior
of the electron antineutrino ($\bar{\nu}_e$) flux during the accretion phase.
The neutronization burst corresponds to the initial $\sim 20$--$30$ ms phase after core bounce, characterized by a sharp spike in the electron neutrino ($\nu_e$) flux produced via electron capture processes. Owing to the suppression of $\bar{\nu}_e$ and heavy-flavor neutrinos ($\nu_x$), this phase provides a clean and robust probe of the neutrino mass ordering through the appearance (IO) or disappearance (NO) of the $\nu_e$ burst peak. 
During the subsequent accretion phase, sustained matter infall leads to the production of all neutrino flavors, with the $\nu_x$ luminosity rising faster than that of $\bar{\nu}_e$. As a consequence of flavor mixing, the $\bar{\nu}_e$ flux in the IO scenario exhibits a faster rise compared to the NO case, making the rise-time of $\bar{\nu}_e$ events a sensitive, though model-dependent, observable for probing the neutrino mass ordering. Neutrino mass dependent self-induced collective neutrino flavor conversions are generally considered to be suppressed during this accretion phase as the neutrinos propagate through even denser matter background. However, fast collective oscillations may still appear 
at the depth of the SN core and can potentially lead to flavor equalization (FE).
In the presence of FE, the $\bar{\nu}_e$ flux becomes an approximate average of the original $\bar{\nu}_e$ and $\nu_x$ fluxes, resulting in an intermediate rise-time behavior that lies between the NO and IO scenarios. While this FE scenario is prominent during the accretion phase, but remains suppressed in the neutronization burst phase due to high electron degeneracy.

Using realistic neutrino fluxes from the Garching CCSN simulations for multiple
progenitor masses 
and incorporating neutrino oscillation scenarios  including
NO, IO, and FE,
we have computed the expected event rates at the detectors using
\texttt{SNOwGLoBES}. A detailed statistical analysis was then performed to
quantify the discrimination power of these observables. Our results show that the neutronization burst remains a robust and largely
model-independent probe of the neutrino mass ordering. In particular, the
appearance (disappearance) of the $\nu_e$ burst peak for IO (NO) provides a
clear and distinctive signature. We find that DUNE exhibits excellent
sensitivity, achieving $\gtrsim 6\sigma$ discrimination for all progenitor
models, with HK also providing competitive sensitivity, $\gtrsim 4\sigma$. This confirms that the
early $\nu_e$ signal is one of the most reliable indicators of the true MO.

In contrast, a similar bin by bin event analysis of the $\bar{\nu}_e$ flux rise-time during the accretion phase is found to be sensitive to the progenitor model, leading to 'apparent' 
degeneracies between different mass models and oscillation scenarios. To
address this limitation, we introduced two improved statistical approaches:
(i) a bin-by-bin cumulative event analysis, and (ii) a ratio-based observable
constructed from cumulative events at characteristic timescales (20 ms and
100 ms). We find that the ratio observable significantly enhances the
separation between MO scenarios by reducing statistical fluctuations and
capturing the intrinsic temporal hierarchy of neutrino emission. In HK, this
method achieves near-complete discrimination even at high confidence levels ($5\sigma$ for $(91-96) \%$ of the models),
while JUNO shows moderate sensitivity ($5\sigma$ for $(52-62) \%$ of the models) limited by its smaller statistics.

Overall, our study highlights the complementary roles of different detectors
and observables in resolving the neutrino mass ordering. While the
neutronization burst offers a clean and model-independent probe, the
accretion phase provides additional sensitivity when analyzed with optimized
statistical tools. The combined analysis of these phases, across multiple
detectors, will be crucial for a definitive determination of the neutrino
mass ordering in the next Galactic supernova event.

Future improvements in supernova simulations, detector systematics, and
understanding of collective neutrino oscillations will further refine these
predictions. Nevertheless, our results demonstrate that the upcoming
generation of neutrino detectors is well equipped to address one of the most
fundamental open questions in neutrino physics.


\acknowledgments
The authors thank Irene Tamborra, Alessandro Mirizzi, and Xilu Wang for useful comments.
The work of P.S. is supported by the National Key R\&D Program
of China (2021YFA0718500), National Natural Science
Foundation of China (Grant Nos. 1252100, 12494570,
12494574), the Chinese Academy of Sciences (Grant
No. E329A6M1) and China’s Space Origins Exploration Program. S.C acknowledges funding received from DST/SERB projects
CRG/2021/002961 and MTR/2021/000540.

\bibliographystyle{JHEP}
\bibliography{biblio}












\end{document}